\documentclass[aps,pra,preprintnumbers,twoside,twocolumn,
superscriptaddress,floatfix,10pt]{revtex4-2}

\usepackage[version=3]{mhchem} 
\usepackage{graphicx}

\usepackage{mathtools}
\usepackage{mathrsfs}
\usepackage{txfonts}
\usepackage{upgreek}
\usepackage{nameref,siunitx}
\usepackage{xr-hyper}
\usepackage[colorlinks=true,citecolor=blue,urlcolor=red,linkcolor=red,filecolor=cyan]{hyperref}
\usepackage{tensor}
\usepackage[T1]{fontenc}

\usepackage{natbib}
\usepackage{cleveref}
\usepackage[mathscr]{eucal}
\usepackage{ulem}
\setcitestyle{super}
\usepackage{enumerate}
\hypersetup{pdfauthor={Name}}

\newcommand{\p}{\partial}
\newcommand{\bn}{\boldsymbol{\nabla}}
\renewcommand{\bn}{\nabla}
\newcommand{\xcomp}{x}
\newcommand{\x}{\mathbf{\xcomp}}
\newcommand{\massdens}{\upvarrho}
\newcommand{\chargedens}{\varrho}
\newcommand{\chempot}{\upmu}
\newcommand{\elpot}{\Phi}
\renewcommand{\elpot}{\varPhi}
\newcommand{\conc}{c}
\newcommand{\freedens}{\upvarphi_\textnormal{H}}

\newcommand{\tempr}{T} 
\newcommand{\pmv}{\upnu}
\newcommand{\diel}{\varepsilon_0}
\newcommand{\dielrel}{\varepsilon_{\ce{R}}}
\newcommand{\dielrelop}{\hat{\varepsilon}_{\ce{R}}}

\newcommand{\satindex}{\alpha^{\ast}}
\newcommand{\tn}{\textnormal{N}}
\newcommand{\Fint}{F^{\ce{int}}}
\newcommand{\Vint}{\mathscr{V}^0}
\newcommand{\lint}{\ell_{\ce{int}}}
\newcommand{\Eel}{E_{\ce{el}}}
\newcommand{\Eth}{E_{\ce{th}}}
\newcommand{\Wroot}{\mathscr{W}}

\newcommand{\ionsize}{a}

\let\kelvin\relax
\let\volt\relax
\let\nano\relax
\let\meter\relax
\let\milli\relax
\let\electronvolt\relax
\begin{document}

\author{Max Schammer} \affiliation{German Aerospace Center,
  Pfaffenwaldring 38-40, 70569 Stuttgart, Germany}
\affiliation{Helmholtz Institute Ulm, Helmholtzstra{\ss}e 11, 89081
  Ulm, Germany}
	
\author{Arnulf Latz} \email{arnulf.latz@dlr.de} \affiliation{German Aerospace Center,
  Pfaffenwaldring 38-40, 70569 Stuttgart, Germany}
\affiliation{Helmholtz Institute Ulm, Helmholtzstra{\ss}e 11, 89081
  Ulm, Germany} \affiliation{Universit\"at Ulm, Albert-Einstein-Allee
  47, 89081 Ulm, Germany}

\author{Birger Horstmann} \email{birger.horstmann@dlr.de}
\affiliation{German Aerospace Center, Pfaffenwaldring 38-40, 70569
  Stuttgart, Germany} \affiliation{Helmholtz Institute Ulm,
  Helmholtzstra{\ss}e 11, 89081 Ulm, Germany}
\affiliation{Universit\"at Ulm, Albert-Einstein-Allee 47, 89081 Ulm,
  Germany}

\title{The Role of Energy Scales for the Structure of Ionic Liquids at
  Electrified Interfaces - A Theory-Based Approach}

\allowdisplaybreaks

\begin{abstract}
  Ionic liquids offer unique bulk and interfacial characteristics as
  battery electrolytes. Our continuum approach naturally describes the
  electrolyte on a macroscale. An integral formulation for the
  molecular repulsion, which can be quantitatively determined by both
  experimental and theoretical methods, models the electrolyte on the
  nanoscale. In this article, we perform a systematic series expansion
  of this integral formulation, derive a description of chemical
  potentials in terms of higher-order concentration gradients, and
  rationalize the appearance of fourth-order derivative-operators in
  modified Poisson equations, recently proposed in this context. In
  this way, we formulate a rigorous multi-scale methodology from
  atomistic quantum chemistry calculations to phenomenological continuum  models. We apply our generalized framework to ionic liquids near  electrified interfaces and perform analytic asymptotic
  analysis. Three energy scales describing electrostatic forces
  between ions, molecular repulsion, and thermal motion determine the
  shape and width of the long-ranging charged double layer. We
  classify the charge screening mechanisms dependent on the system
  parameters dielectricity, ion size, interaction strength, and
  temperature. We find that the charge density of electrochemical
  double layers in ionic liquids either decays exponentially, for
  negligible molecular repulsion, or oscillates continuously. Charge
  ordering across several ion-diameters occurs if the repulsion
  between molecules is comparable with thermal energy and Coulomb
  interaction. Eventually, phase separation of the bulk electrolyte
  into ionic layers emerges once the molecular repulsion becomes
  dominant. Our framework predicts the exact phase boundaries among
  these three phases as a function of temperature, dielectricity and
  ion-sizes.
\end{abstract}

\maketitle

\section{Introduction}
\label{sec:introduction-1}
	
Strong electrostatic correlations in crowded environments play an
important role in biology, chemistry and
physics.\cite{eisenberg2013interacting,grosberg2002colloquium,henderson2009attractive}
For example, in molecular biology, they account for DNA
packing,\cite{chen2006local} which is crucial for the compactification
of genetic materials in viruses,\cite{bloomfield1991condensation}
impact the cytoskeleton organization,\cite{wong2010electrostatics} and
influence transport in ion channels.\cite{eisenberg2010energy}
Furthermore, such correlations explain the thermodynamic stability of
plasmas,\cite{fisher1993criticality,levin1996criticality} and charged
colloidal suspensions.\cite{van1997van,hansen2000effective}
	
Surprisingly, the complexity of these phenomena can be un\-der\-stood
to a large degree by models derived initially for electrolyte
solutions.\cite{levin2002electrostatic} Starting from the fundamental
Debye-H\"uckel theory for dilute solutions,\cite{huckel1924theorie}
increasingly accurate models for concentrated electrolytes were
developed,\cite{PhysRevE.75.021502} taking more complex Coulomb
correlations into account.
	
Because ionic liquids (ILs) consist only of positive and negative ions
without neutral solvent, they constitute the extreme limit for the
examination of electrostatic correlations in electrolytic
solutions. Indeed, ILs possess characteristic properties in the
bulk-regime,\cite{doi:10.1021/cr500411q,doi:10.1021/acs.jpcb.7b01654}
but also near electrified interfaces.\cite{fedorov2014ionic} This
makes them highly attractive from both fundamental and applied
perspectives.\cite{plechkova2008applications,B906273D,torimoto2010new,werner2010ionic,welton1999room,endres2004ionic,armand2011ionic}
The study of interfacial electrochemistry is of wide-ranging
interest. For example, the behavior of ILs near electrified interfaces
has paramount importance for their performance as battery
electrolytes.\cite{endres2010solvation,macfarlane2010ionic}
	
Theoretical studies of ILs near electrified interfaces discuss the
structure of charged electrochemical double layers (EDL) on
atomistic/molecular scales. These include classical density functional
theory (cDFT) simulations, and molecular-dynamics (MD)
simulations. cDFT gives detailed insights into the arrangement of
molecules in the
EDL.\cite{wu2011classical,henderson2012electrochemical,jiang2014time}
MD resolves the molecular motion and can elucidate the EDL
structure.\cite{sharma2016structure,sharma2015structure,hu2013molecular}
	
However, cDFT/MD simulations are limited by their com\-pu\-ta\-tio\-nal
costs. Simulations at length-scales above the nanometer scale are
hardly accessible to the ato\-mis\-tic / mo\-lecular approach. Thus,
continuum theories, and mean\--field\--theo\-ries (MFT) provide a
complementary methodology for the simulation of larger systems, where
the microscopic details can be neglected, or are used as averaged
parameters (\textit{e.g.} constant dielectric parameters).
	
Usually, MFTs for electrolytes are based on lattice gas models of
ions, first proposed by Bikermann.\cite{doi:10.1080/14786444208520813}
Recently, MFTs have attracted great interest fot the study for
ILs. As proposed by Santangelo for aqueous
systems,\cite{santangelo2006computing} the extension of MFTs by
higher-order electrostatic correlations is useful for the
description of long-ranged structures emerging in
electrolytes. Bazant, Storey and Kornyshev (BSK) applied this approach
to ILs near electrified interfaces.\cite{bazant} By using a
phenomenological model, which is based on a generalized
Ginzburg-Landau functional, BSK describe charge oscillations known as
overscreening and charge saturation known as crowding. Yochelis et
al. rationalized this approach, and extended it to bulk
properties.\cite{yochelis2014transition,yochelis2014spatial,doi:10.1021/acs.chemmater.5b00780,doi:10.1021/acs.chemmater.5b00780,doi:10.1021/acs.jpclett.6b00370,PhysRevE.95.060201}
However, MFT models are usually restricted to equilibrium effects of
binary ILs with structureless bulk, although rare MFT models,
complemented by continuum methods
\cite{PhysRevLett.115.106101,doi:10.1021/acs.jpclett.6b00370} and
extended to the ternary case,\cite{doi:10.1021/acs.jpclett.7b03048}
exist.

This highlights the advantage of continuum frameworks, which describe
dynamical transport processes. In addition, continuum models based on
rigorous physical assumptions\cite{bothe2015continuum} identify
coupled phenomena arising from the interplay of mechanics,
thermodynamics, and electromagnetic
theory.\cite{kovetz2000electromagnetic} Furthermore, this approach
allows to develop a unified, thermodynamically consistent framework that provides the common theoretical basis for the description of
different electrochemical
systems.\cite{schammer2020theory,latz2015multiscale,schmitt2019zinc,becker2020,single2019theory}
Continuum models are not restricted to binary or ternary systems, because
they can be formulated for arbitrary many species, charged and
uncharged. Thus, they apply to more realistic electrolytes.
	
Recently, we proposed such a novel continuum transport
theory.\cite{schammer2020theory} In this theory, we take account for
steric effects via the mean-volume, which is due to finite molar
volumes of the ion species. For this purpose, we impose a volume constraint on the electrolyte. This mechanism
stabilizes the bulk structure against Coulomb
collapse
\cite{hansen2006theory} and leads to charge-saturation near
electrified interfaces. Thus our theory resolves the deficiencies of
the classical Poisson-Boltzmann (PB) theory, which predicts
unrealistically high
interface-concentrations.\cite{PhysRevE.75.021502} Furthermore, the existence of finite molar volumes of the electrolyte species leads to a pressure dependence of the chemical potentials (in accordance with thermodynamic arguments).\cite{Landstorfer2016} 
	
However, this bulk-framework cannot describe the emergence of
long-range structures in ILs near electrified inter\-fa\-ces. Therefore,
in a joint experimental / theoretical work, we extended our framework
with non-local interactions and validated it with results obtained
from atomic-force-mi\-cros\-copy.\cite{C7CP08243F} Thus, we extended the
mean-volume-effect of the bulk theory with molecular volume exclusion
due to hardcore re\-pul\-sion.  Our holistic framework allows us to couple
dynamic transport processes occurring in the bulk-electrolyte with
inter\-fa\-cial electrochemical processes. Thus, we provide a continuum
model which bridges the length-scales from nano-meters, \textit{e.g.},
EDL, to millimeters, \textit{e.g.}, battery cells. Moreover, our
framework allows us to connect the continuum description with correlation
functions generated by MD.
	
However, the dependence of EDL structures on molecular repulsion,
molecular size, temperature, and dielectricity is still unknown. In
this paper, we derive such an understanding with asymptotic
analysis. To this aim, we present our thermodynamically consistent
transport theory with an integral formulation of non-local
interactions in \cref{sec:free-energy-funct}. These correlations
represent atomistic volume exclusion and lead to modified constitutive
equations
(\cref*{eq:SI_const_eq_entro,eq:SI_const_eq_elf,eq:SI_const_eq_magf,eq:SI_const_eq_chempot}). Moreover,
the interactions impose contributions to the stress-tensor, and thus
modify the mechanical coupling to the transport equations
(\cref*{eq:SI_const_eq_stress_tensor}). In
\cref{sec:gradient-expansion}, we approximate the interaction
functional with a gradient expansion, which facilitates the analytic
asymptotic analysis of the EDL structure. In \cref{sec:binary-il}, we
apply our extended framework to study the EDL structure for neat
ILs. When we non-dimensionalize our dynamical description in
\cref{sec:energy-scal-dimens}, three competing energy scales
describing electrostatic forces among ions, molecular repulsion, and
thermal motion appear in the theory. Because our focus lies on the
formation of equilibrium structures, we discuss the stationary state
in \cref{sec:stationary-state}. In \cref{sec:small-large-potent}, we
discuss limiting cases of our stationary theory.
	
We perform numerical simulations and analytic asymptotic analysis to
study the interplay and the effect of the competing energy scales on
EDL structures. First, in \cref{sec:underscreening}, we discuss the
EDL structure for the mean volume constraint. Second, in
\cref{sec:inter-symm-ions}, we incorporate molecular repulsion into
our analysis and classify the EDL structure dependence on the relation
between competing energy scales.

\section{Theory}
\label{sec:theory}
	
\subsection{Generalized Transport Theory}
\label{sec:free-energy-funct}
	
Recently, we have proposed a free energy functional
$F^{\ce{b}} = \int\ce{d}V\, \massdens\freedens$ for the dynamical
description of ionic liquids in the
bulk-phase.\cite{schammer2020theory} In this bulk-model, the Helmholtz
free energy density
$\freedens(\varUpsilon)=\freedens( \tempr, \conc_1, \ldots,
\conc_\tn,\boldsymbol{D}, \boldsymbol{B}, \boldsymbol{\upkappa})$ is a
function of the variables temperature $T$, concentrations $c_\alpha$,
dielectric displacement $\boldsymbol{D}$, magnetic field
$ \boldsymbol{B}$, and strain-rate tensor
$\boldsymbol{\upkappa}$. This variable-set $\varUpsilon$ constitutes
material-specific properties of multi-component, viscous and
polarizable media in the liquid state.
	
In contrast, models describing non-local interactions rely on
functionals $\Fint[\varUpsilon]$, such that the free energy takes the
form
\begin{equation}
  \label{eq:interaction_extension}
  F[\varUpsilon] = \Fint[\varUpsilon]+  F^{\ce{b}}(\varUpsilon) 
  = \Fint[\varUpsilon] +  \int \ce{dV}\, \massdens\freedens .
\end{equation}

This functional approach constitutes a more general description for
electrolyte materials and allows the incorporation of non-local
correlations between field quantities. Such correlations typically
arise from microscopic effects occurring on the nano-meter scales,
e.g., in the vicinity of electrified interfaces. Despite the conceptual
difference between the functional approach
\cref{eq:interaction_extension} and the canonical bulk approach, the
derivation of the resulting transport theory is rather similar to the
rationale outlined in great detail in
Ref.~\citenum{schammer2020theory} for the free energy
$F^{\ce{b}}=\int\ce{dV}\massdens\freedens$. We present a detailed
derivation for the functional approach in the Supporting Information (see
\cref*{sec:SItheory}).

The extension of the free energy according to
\cref{eq:interaction_extension} leads to modified constitutive
equations for entropy density $s$, electric field strength
$\boldsymbol{\mathscr{E}} $, magnetic field
$\boldsymbol{\mathscr{H}} $, chemical potentials $\chempot_\alpha$,
and the stress tensor $ \boldsymbol{\upsigma}$ in the form of
functional-derivatives (see
\cref*{eq:SI_const_eq_entro,eq:SI_const_eq_elf,eq:SI_const_eq_magf,eq:SI_const_eq_chempot,eq:SI_const_eq_stress_tensor}
in \cref*{sec:SItransport-theory}). We evaluate this framework for the
bulk-energy of a linear dielectric medium discussed in
Ref.~\citenum{schammer2020theory}, see
\cref*{eq:SI_model_bulk_energy}. The resulting forces are supplemented
by contributions stemming from the non-local correlations (\cref*{eq:SI_chemical_force}).

For the remaining part of this work, we neglect thermal driving forces
by setting the temperature equal to constant values, and assume the
electrostatic limit, $\boldsymbol{B}=0$ and
$\boldsymbol{\mathscr{H}}=0$. This determines the electric field
$\boldsymbol{E}=\boldsymbol{\mathscr{E}}$ by the electrostatic
potential, $\boldsymbol{E}=-\bn\elpot$.
	
\subsection{Gradient Expansion of Molecular Interactions}
\label{sec:gradient-expansion}
	
We proposed a model for
hardcore-interactions based on a convolution-functional for the
interaction free energy in Ref. \citenum{C7CP08243F},
\begin{equation}
  \label{eq:F_int_full}
  \Fint
  =
  \frac{1}{2}
  \sum_{\alpha,\,\beta}^{\tn}
  \iint \ce{d}\xcomp^3 \ce{d} y^3 \
  \mathscr{F}_{\alpha\beta}(|\x-\mathbf{y}|)
  \conc_\alpha(\x)
  \conc_\beta(\mathbf{y}),
\end{equation}
leading to transport contributions in the form of (see \cref*{sec:SIfunct-deriv})
\begin{equation}
  \label{eq:functderiv_integralstyle}
  \frac{\updelta\Fint}{\updelta\conc_\alpha}(\x)
  =
  \sum_{\beta}^{\tn}
  \int \ce{d} y^3  \
  \mathscr{F}_{\alpha\beta}(\lvert\x- \mathbf{y}\rvert)
  \conc_\beta(\mathbf{y}).
\end{equation}
The symmetric potentials $\mathscr{F}_{\alpha\beta}$ determine the
correlation length $\lint$, and the magnitude of the interaction. The number of additional parameters describing this interaction
depends upon the model for $\mathscr{F}_{\alpha\beta}$. In a previous
publication we used a Lennard-Jones-type force-field for
$\mathscr{F}_{\alpha\beta}$.\cite{C7CP08243F} Such potentials are
often used in the
literature.\cite{bedrov2019molecular,heinz2013thermodynamically,eisenberg2011mathematical,
  lin2014new, lin2015multiple} Furthermore, because
\begin{equation}
  \label{eq:rel_intpot_corrltn_fct}
  \frac{\updelta^2\Fint}{\updelta\conc_{\beta}(\mathbf{z})\updelta\conc_{\alpha}(\mathbf{x})}
  = \mathscr{F}_{\alpha\beta}(\lvert\x-\mathbf{z}\rvert),
\end{equation}
the potentials $ \mathscr{F}_{\alpha\beta}$ determine the direct pair
correlation functions used in liquid state
theory.\cite{hansen2006theory}

Experimental results suggest that such interactions typically decay
after some ionic diameters.\cite{C7CP08243F} Thus, we focus on
potentials $\mathscr{F}_{\alpha\beta}$ ranging over the size of one
molecule. Their extend $\lint$ is large compared to the exponential
decay of the electric field, \textit{i.e.}, the
Debye-length,\cite{Lee2017,Rotenberg_2018} yet small compared to the
battery cell.
	
In the SI (\cref*{sec:SIgradient-expansion}), we show that such a 
con\-vo\-lu\-tion functional $\Fint$ can be approximated in
power-series of concentration gradients when
$\mathscr{F}_{\alpha\beta}$ are short-ranged,
\begin{equation}
  \label{eq:fint_grad_exp}
  \Fint[\conc_\gamma]
  =
  \frac{1}{2}
  \sum_{\alpha,\,\beta}^{\tn}
  \sum_{n=0}^\infty
  \varGamma^{2n}_{\alpha\beta} \int \ce{d}^{3} y \
  \conc_\alpha(y) \cdot  \bn^{2n} \conc_\beta(y),
\end{equation}
where
\begin{equation}
  \label{eq:expansioncoefficients}
  \varGamma^n_{\alpha\beta}(\mathscr{F}_{\alpha\beta})
  = \frac{1}{n!} \cdot \int
  \ce{d}^{3}\xcomp\, \mathscr{F}_{\alpha\beta}(|\xcomp|) \cdot \xcomp^n. 
\end{equation}
Here, $\varGamma^n_{\alpha\beta}$ are symmetric perturbation
coefficients of dimension
$[\varGamma^{2n}_{\alpha\beta}] =
\si{\joule\meter}{}^{3+2n}\si{\per\mol\squared}$. We state the
complete free energy functional for IL electrolytes in the SI (\cref*{eq:SI_const_eq_stress_tensor}).

The excess chemical potentials are determined via functional
derivatives (\cref*{eq:SI_const_eq_chempot}). In
\cref*{sec:SIfunct-deriv}, we show that this Ansatz leads to force contributions  
\begin{equation}
  \label{eq:chempot_grad_exp}
  \frac{\updelta \Fint}{\updelta \conc_\alpha(z)}
  = \sum_{\beta=1}^{\tn}
  \sum_{n=0}^{\infty} \varGamma^{2n}_{\alpha\beta} \cdot \bn^{2n}
  \conc_\beta(z).
\end{equation}
The corresponding electrochemical potentials for ionic liquid
electrolytes are
\begin{multline}
  \label{eq:elpot_force}
  \bn \chempot^{\ce{el}}_\alpha = \left( F z_\alpha - \pmv_\alpha
    \chargedens\right) \bn\elpot + R \tempr \cdot (\bn
  \conc_\alpha)/\conc_\alpha - R \tempr \pmv_\alpha \bn \conc
  \\
  + \sum_{n=0}^\infty \sum_{\beta,\,\gamma}^{\tn} \left(
    \updelta^{\alpha}_{\beta} - \pmv_\alpha \conc_\beta \right)
  \varGamma^{2n}_{\beta\gamma} \cdot \bn^{2n+1} \conc_\gamma.
\end{multline}
	
We specify our electrolyte model and assume a one-di\-men\-sio\-nal
Gaussian interaction potential for symmetric ions,
\begin{equation}
  \label{eq:gauss_shaped_interaction}
  \mathscr{F}_{\alpha\beta}
  = \left(2\sqrt{2\pi}\right)^3\cdot
  \Vint \cdot (N_A)^2 \cdot 
  \exp\left[
    -2\left(\frac{\xcomp\pi}{\ionsize}\right)^2
  \right].
\end{equation}
Here, $\Vint$ denotes the characteristic interaction energy, and
$\ionsize$ is the extension of ion-pairs. This material parameter
determines the correlation length $\lint=\ionsize/\pi\sqrt{2}$ of the
interaction. We assume that $\ionsize$ emerges naturally from the
common molar volume $\pmv=\pmv_++\pmv_-$ via $\pmv=N_A\ionsize^3$,
which is justified below (see \cref{sec:interactions_small-potentials} and \cref{eq:terminal_k}). Thus, only $\Vint$ is introduced as a novel
independent material parameter. In contrast, potentials of Lennard
Jones type need at least one more parameter for the well depth. In the
case of a binary system, the lowest order expansion coefficients (\cref{eq:expansioncoefficients}) of the inter-species correlations for
the Gaussian model in \cref{eq:gauss_shaped_interaction} are (\cref*{sec:SIgauss-model-inter})
\begin{gather}
  \label{eq:zero_order_gamma}
  \varGamma^0_{12} = \Vint N_A \pmv = \Vint \left(N_A \right)^2
  \ionsize^3,
  \\
  \label{eq:second_order_gamma}
  \varGamma^{2}_{12} = \frac{\lint^2}{2} \cdot \varGamma^0_{12} =
  \frac{\ionsize^2}{4\pi^2} \cdot \varGamma^0_{12}.
\end{gather}

\subsection{Binary IL}
\label{sec:binary-il}
In this section, we apply our formalism to binary ILs at electrified
interfaces. Thus, we use the extended electrochemical forces
\cref{eq:elpot_force} in our multicomponent framework derived in
Ref. \citenum{schammer2020theory}.

\begin{figure}[!htb]
  \includegraphics[width=.6\columnwidth]{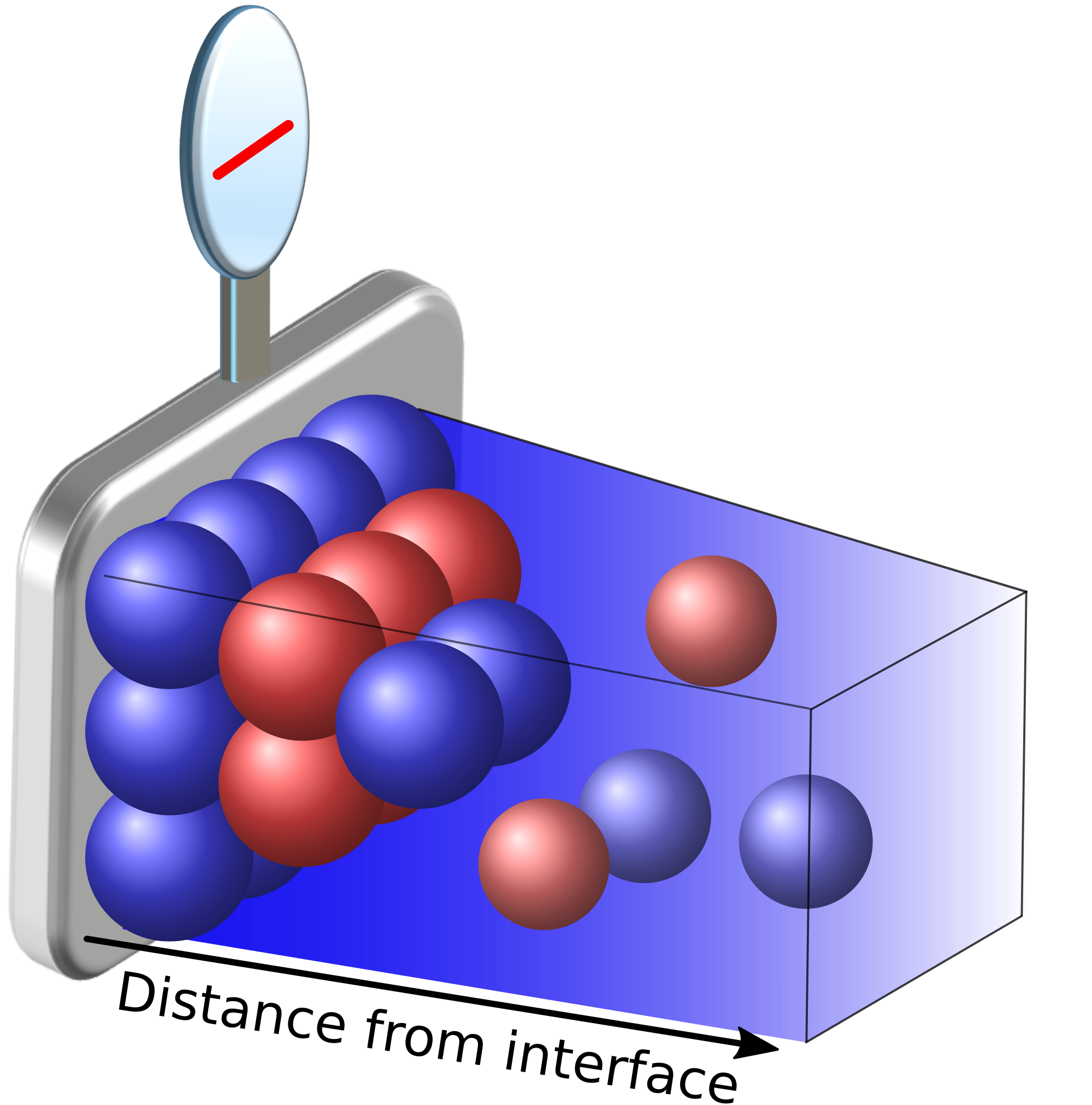}
  \caption{Scheme of the as-modelled set-up. The binary IL-electrolyte
    \ce{PYR[1,4]TFSI} is subject to the negatively charged interface
    at the left, which causes the formation of an electrochemical
    double layer (EDL). Charge-ordering diminishes with increasing
    distance from the interface (towards the right side), and the
    electrolyte is electroneutral in the bulk.}
	\label{fig:scheme_interface}
\end{figure}
	
As discussed in \cref{sec:free-energy-funct}, binary electrolytes are
described with the variables electric potential $\elpot$, charge
density $\chargedens$, and center-of-mass convection velocity
$\mathbf{v}$. Furthermore, the electric conductivity $\kappa$ is the
only independent transport parameter in this
case.\cite{schammer2020theory} The dynamical transport equations are
\begin{gather}
  \label{eq:charge_eq}
  \partial_t \chargedens = - \bn \left( \chargedens\mathbf{v}\right) -
  \bn \boldsymbol{\mathcal{J}},
  \\
  \label{eq:poisson_eq_simple}
  \chargedens = - \diel \bn \dielrel \bn \elpot,
  \\
  \label{eq:vel_eq}
  \bn\mathbf{v} = ( M_+\pmv_- - M_-\pmv_+) / F z_+M_{\ce{IL}} \cdot
  \bn \boldsymbol{\mathcal{J}}.
\end{gather}
Here, $M_\pm$ are the molar masses of the ionic species which sum to
$M_{\ce{IL}}$, and $\boldsymbol{\mathcal{J}}$ is the electric current
relative to the center-of-mass motion,
$ \boldsymbol{\mathcal{J}} = \kappa( M_+ \cdot \bn\chempot^{\ce{el}}_-
- M_- \bn\chempot^{\ce{el}}_+ )/M_{\ce{IL}}Fz_+$.

Solutions to \cref{eq:charge_eq,eq:poisson_eq_simple,eq:vel_eq}
determine the ionic con\-cen\-tra\-tions via
$\chargedens = Fz_+ (\conc_+ - \conc_-)$ (charge-conservation), and
via the Euler equation for the volume \cref*{eq:SI_elyte_eq_of_state}.

We restrict our setup to one spatial dimension and assume that the
inert electrified interface is located at $\xcomp=\num{0}$. The
electroneutral boundary condition
$\chargedens(\xcomp{\to}\infty)=\num{0}$ implies that the bulk
concentration $\conc^{\ce{b}}=\conc_{\pm}(\xcomp{\to}\infty)$ is
completely determined by the total partial molar volume
$\pmv=\pmv_++\pmv_-$ via $\conc^{\ce{b}}\cdot\pmv = \num{1}$. Because
binary ILs are electrically neutral, $z_-=-z_+$, and we choose
$z_+>0$.

We neglect viscous forces in our discussion of the EDL
($\bn\boldsymbol{\tauup}=\num{0}$). Therefore, the Gibbs-Duhem
relation (see \cref*{eq:SI_trivial_gibbs_duhem}) becomes
$\conc_+\bn
\chempot^{\ce{el}}_++\conc_-\bn\chempot_-^{\ce{el}}=\num{0}$, and the
expression for the electric flux simplifies to
$ \boldsymbol{\mathcal{J}} = -\kappa\pmv\massdens/Fz_+M_{\ce{IL}}
\cdot \bn \chempot_{\ce{IL}}$, where we use the chemical potential of
the anion-species to determine the IL-electrolyte
($\bn \chempot_{\ce{IL}} = \bn \chempot^{\ce{el}}_-$), 
\begin{multline}
  \label{eq:chempot_IL}
  \bn \chempot_{\ce{IL}} 
  = \bn \left(
  Fz_+ \elpot
    - 
      \gammaup_+ \frac{\updelta\Fint}{\updelta\conc_-}
      + \gammaup_- \frac{\updelta\Fint}{\updelta\conc_+}
      - \right. \\ \left.
    - RT \gammaup_+\ln\left[
      \frac{\conc_-}{\conc^{\ce{b}}}\right]
     + RT \gammaup_-\ln\left[
        \frac{\conc_+}{\conc^{\ce{b}}}\right]
    \right).
\end{multline}
Here, we introduced the relative magnitude of the molar volumes
$\gammaup_\pm = \pmv_\pm/\pmv$. Thus, the forces given by
\cref{eq:chempot_IL} depend upon the model for $\Fint$. Furthermore,
by using the Gauss-model (see \cref{eq:gauss_shaped_interaction}), we can
either close the forces via the ``complete'' integral equation
(\cref{eq:functderiv_integralstyle}) or we can use the gradient expansion
(\cref{eq:chempot_grad_exp}).

We want to apply the half-cell potential $\varDelta\phi$. Because the
electric potential $\elpot$ is continuous across the
electrode-electrolyte interface, $\elpot(0)$ in the electrolyte is
subject to the boundary condition
\begin{equation}
  \label{eq:BC_phi}
  \elpot(0) - \elpot(\xcomp\to\infty) = \varDelta\phi.
\end{equation}
Without loss of generality, we set the electrolyte potential in the
bulk to zero, $\lim_{\xcomp\to\infty}\elpot=\num{0}$. Hence,
$\varDelta\phi=\elpot(\num{0})$ is the potential applied to the
electrode.

We perform one-dimensional numerical simulations of this system of
equations \ref{eq:charge_eq}, and \ref{eq:poisson_eq_simple} in the completely
dissociated state, subject to an inert electrified interface (for more
details, see \cref*{sec:comp-deta}). This electrolyte is part of the
IL-family composed of \ce{TFSI} anions and \ce{PYR} cations. Because of
their excellent electrochemical properties, these ILs are widely
studied and used for applications in lithium-ion
batteries.\cite{karuppasamy2020ionic} We state the electrolyte
parameters in the SI (see \cref*{sec:SI_binary-ionic-liquid}).

\subsection{Energy Scales and Dimensions}
\label{sec:energy-scal-dimens}
In this section, we clarify the notation and state the non-dimensional
form of principal
quantities appearing in our theory. For a complete discussion,
we refer to the SI (see \cref*{sec:SInonimensionalization}), where we
share the motivation for our choices.

We introduce dimensionless variables for electric potential
$\tilde{\elpot} = \elpot\cdot Fz_+/RT$, charge density $\tilde{\chargedens} = \chargedens \cdot \pmv\tilde{\conc}^{\ce{b}}/Fz_+$, and
concentration $\tilde{\conc}_\alpha = \conc_\alpha \cdot \pmv\tilde{\conc}^{\ce{b}}$.  As a consequence, the Euler
equation for the volume becomes
$\tilde{\conc}^{\ce{b}} =
\gammaup_+\tilde{\conc}_++\gammaup_-\tilde{\conc}_-$,
with the dimensionless molar volumes $\gammaup_\pm = \pmv_\pm/\pmv$ and the dimensionless bulk concentration
$\tilde{\conc}^{\ce{b}}$. The Poisson equation suggests defining the generalized
Debye-length
\begin{equation}
  \label{eq:def_debye_length}
  L_{\ce{D}}
  = \sqrt{
    \frac{k_{\ce{B}}\tempr\diel\dielrel\ionsize^3\tilde{\conc}^{\ce{b}}}{(ez_+)^2}
  }.
\end{equation}
Here, we used $F=eN_{\ce{A}}$ for the Faraday constant, where $e$ is
the elementary charge, and $N_{\ce{A}}$ is Avogadro's constant, and
the model $\pmv=N_{\ce{A}}\ionsize^3$ for the partial molar volumes
introduced above. This Debye-length differs from the canonical
definition by the asymmetry-factors
$\gammaup_\pm$.\cite{schmickler2010interfacial} However, it reproduces
the textbook definition for symmetric ions
($\gammaup_\pm=\num{0.5}$). $L_{\ce{D}}$ becomes minimal for
$\gammaup_\pm=\num{0.5}$ because the mixing entropy of a binary
electrolyte is extremal for equal ion-size. Thus, asymmetry increases
the Debye-screening length.

With this length scale, we non-dimensionalize our grid, \textit{viz.} $\tilde{x} = x/L_{\ce{D}}$ and $\tilde{\bn} = L_{\ce{D}} \cdot \bn $, and obtain the dimensionless  Poisson-equation, 
\begin{equation}
  \label{eq:non_dim_trivial_poisson}
  \tilde{\chargedens} = - \tilde{\bn}^2\tilde{\elpot}.
\end{equation}

In the SI (see \cref*{sec:SInonimensionalization}), we
non-dimensionalize the transport equations (see
\cref{eq:charge_eq,eq:vel_eq}) for binary symmetric ILs. Because we
neglect convective effects in our EDL discussion, the complete set of
equations consists of the Poisson equation and one transport equation
for the charges.  By substituting
\cref{eq:functderiv_integralstyle} into \cref{eq:chempot_IL}, we find
for the integral description
\begin{multline}
  \label{eq:dyn_tt_integral}
  \p_{\tilde{t}}\tilde{\chargedens}
  = \tilde{\bn}\Biggl[
  \left(1 + \chiup\tilde{\chargedens}\right) \cdot \tilde{\bn}
    \Biggl(\tilde{\elpot}
    -  \gammaup_+
    \ln\left[
      \frac{\tilde{\conc}_-}{\tilde{\conc}^{\ce{b}}}\right]
      + \gammaup_-
      \ln\left[
        \frac{\tilde{\conc}_+}{\tilde{\conc}^{\ce{b}}}\right]
      -
    \\
    - \int\ce{d}\tilde{\xcomp}^3
    \tilde{\mathscr{F}}_{\alpha\beta}(|\tilde{\x}-\tilde{\mathbf{y}}|) 
    \tilde{\chargedens}(\tilde{\mathbf{y}})
    \Biggr)
  \Biggr],
\end{multline}
where
$\chiup = (\gammaup_-
M_+/M_{\ce{IL}}-\gammaup_+M_-/M_{\ce{IL}})/\tilde{\conc}^{\ce{b}}$
measures the ``asymmetry'' of the ion-species, and $\p_{\tilde{t}} = \diel\dielrel/\kappa \cdot \p_{t} $.

The interaction potential is non\--di\-men\-sio\-na\-lized  (see
\cref*{eq:SI_dimless_interaction_pot}) by two energy-scales for
thermal energy $\Eth$ and electrostatic energy $\Eel$,
\begin{gather}
  \label{eq:en_scale_eth}
  \Eth = k_{\ce{B}}\tempr \cdot \frac{ \tilde{\conc}^{\ce{b}}
  }{2\gammaup_+\gammaup_-},
  \\
  \label{eq:en_scale_eel}
  \Eel = \frac{(ez_+)^2}{\num{4}\pi\diel\dielrel} \cdot
  \frac{\num{1}}{\num{4}\gammaup_+\gammaup_-} \cdot
  \frac{\num{1}}{\ionsize},
\end{gather}
such that $\tilde{\mathscr{F}}_{\alpha\beta} = \mathscr{F}_{\alpha\beta}/(N_{\ce{A}})^2\Eth \cdot (L_{\ce{D}}/\ionsize)^3$
In the case of symmetric ions $\gammaup_\pm=\num{0.5}$, these energy
scales take the textbook form for thermal energy and Coulomb energy of
charges at distance $\ionsize$. Apparently, both energy scales are
coupled by the generalized Debye-length $L_{\ce{D}}$,
\begin{equation}
  \label{eq:corrl_eth_and_Eel}
  \Eth/\Eel = 8\pi \left(L_{\ce{D}}/\ionsize\right)^2.
\end{equation}

The integral form \cref{eq:dyn_tt_integral} for the transport equation
allows to relate our continuum framework to MD-simulations as
discussed in \cref{sec:relat-md-meth}.

In this article we restrict the gradient-ex\-pan\-sion of the
interaction to the trivial and first non-trivial modes
($\tilde{\varGamma}^{0}_{+-} = \Vint/\Eth$ and
$\tilde{\varGamma}^{2}_{+-}= 2/\pi \cdot \Vint/\Eth\cdot \Eel/\Eth$,
see \cref*{eq:SI_nondim_pertmodes}) and obtain
\begin{multline}
  \label{eq:dyn_tt_gradexpansion}
  \p_{\tilde{t}}\tilde{\chargedens}
  = \tilde{\bn}\Biggl[
  \left( 1 + \chiup\tilde{\chargedens}\right)
  \cdot  \tilde{\bn} 
  \Biggl(
    \tilde{\elpot}
    -  \gammaup_+
      \ln\left[
        \frac{\tilde{\conc}_-}{\tilde{\conc}^{\ce{b}}}
      \right]
      + \gammaup_-
      \ln\left[
        \frac{\tilde{\conc}_+}{\tilde{\conc}^{\ce{b}}}
      \right]
      \\
      - \frac{\Vint}{\Eth}
      \left( 1 +\frac{2}{\pi}
        \frac{\Eel}{\Eth}
         \cdot\tilde{\bn}^2\right) \tilde{\chargedens}
      \Biggr)
      \Biggr].
\end{multline}

\subsection{Stationary State}
\label{sec:stationary-state}

Because our focus is the formation of equilibrium structures, we
discuss the system of equations in the stationary limit. This allows
us to integrate the differential equations using electroneutral boundary
conditions, which results in a simplified description susceptible to
analytic techniques.

Stationarity ($\partial_t\chargedens{=}\num{0}$) implies that all fluxes
are constant. Here, we have no flux conditions
$\tilde{\boldsymbol{\mathcal{J}}} = \tilde{\mathbf{v}} = \num{0}$,
which implies that both species are in equilibrium, $
\bn\chempot^{\ce{el}}_+ =   \bn\chempot^{\ce{el}}_- $, \textit{i.e.}, 
\begin{equation}
  \label{eq:poteL_is_zero}
   \tilde{\bn}\tilde{\chempot}_{\ce{IL}}= \num{0}.
\end{equation}
		
Thus, the stationary state for the binary electrolyte is described by
the Poisson equation and \cref{eq:poteL_is_zero}. Here, we evaluate
the equilibrium condition using the gradient description
(\cref{eq:chempot_IL}) in the non-dimensionalized form (see
\cref*{eq:SI_nondim_pertmodes} and \cref{eq:dyn_tt_gradexpansion}), and integrate the
result using electroneutral boundary conditions in the bulk,
\begin{multline}
  \label{eq:non_dim_trivial_forcelaw}
  \num{0} = \tilde{\elpot} - \frac{\Vint}{\Eth} \left( \num{1} +
    \frac{\num{2}}{\pi}  \frac{\Eel}{\Eth} \tilde{\bn}^2
  \right) \tilde{\chargedens}- \left( \gammaup_+
    \ln\left[\frac{\tilde{\conc}_-}{\tilde{\conc}^{\ce{b}}}\right]
    - \right. \\ \left. -
    \gammaup_- \ln\left[
      \frac{\tilde{\conc}_+}{\tilde{\conc}^{\ce{b}}}\right] \, 
  \right).
\end{multline}
	  
Apparently, in contrast to the dynamical case where electrolyte
momentum is important, the molar masses appearing as parameters in the
fluxes $\mathbf{v}$ and $\boldsymbol{\mathcal{J}}$ become irrelevant in the
stationary limit. Instead, the relative magnitude of the molar
volumes $\gammaup_\pm$ enters the system of equations. This highlights
the principal role of molar volumes as parameters in the stationary
state and is a consequence of the Euler equation for the volume,
\cref*{eq:SI_elyte_eq_of_state}.
	  
For completeness, we state the integral  transport equation \ref{eq:dyn_tt_integral}
in the stationary limit (see \cref*{sec:inter-contr-chem})
\begin{multline}
  \label{eq:non_dimm_intgrtd_intgrl_tt}
  \num{0} = \tilde{\elpot} - \int\ce{d}\tilde{y}
  \tilde{\mathscr{F}}_{+-}\left(\tilde{\xcomp},\tilde{y}\right)
  \tilde{\chargedens}(\tilde{y}) - \left( \gammaup_+
    \ln\left[\frac{\tilde{\conc}_-}{\tilde{\conc}^{\ce{b}}}\right]
    - \right. \\ \left. -
    \gammaup_- \ln\left[ \frac{\tilde{\conc}_+}{\tilde{\conc}^{\ce{b}}}\right]
  \right).
\end{multline}
	
\subsection{Small and Large Potentials}
\label{sec:small-large-potent}
	
\Cref{eq:non_dim_trivial_poisson} and
\cref{eq:non_dim_trivial_forcelaw} (or
\cref{eq:non_dimm_intgrtd_intgrl_tt}) constitute the complete set of
equations, necessary to describe a binary IL-electrolyte in a stationary
state. In \cref{sec:numer-invest} and \cref{sec:numeric-validation}, we
solve these equations using numerical methods. Our goal is to
supplement these numerical methods by an analytic examination of the
gradient description. However, the analytic solution of the gradient
description is hindered by the higher or\-der gradients appearing in
\cref{eq:non_dim_trivial_forcelaw} and by the different prefactors of
the logarithmic terms (in general,
$\gammaup_+\neq\gammaup_-$). Therefore, we distinguish different
limiting cases in our analysis in
\cref{sec:story-telling,sec:math-analys}. In the SI, we describe the
special case of symmetric ion-species (see
\cref*{sec:integr-transp-equat,sec:SImean-volume-effect}).
	
In \cref{sec:story-telling} we show that the limiting case of
small charge-densities, $\tilde{\chargedens}\ll \num{1}$, is
useful. In this case, we can expand the logarithmic term in
\cref{eq:non_dim_trivial_forcelaw} and
\cref{eq:non_dimm_intgrtd_intgrl_tt} around the electroneutral state,
\begin{equation}
  \label{eq:log_taylor}
  \gammaup_+
  \ln\left[\tilde{\conc}_-/\tilde{\conc}^{\ce{b}}\right]
  - \gammaup_-
  \ln\left[\tilde{\conc}_+/\tilde{\conc}^{\ce{b}}\right]
  \approx -\tilde{\chargedens},
\end{equation}
such that
\cref{eq:non_dim_trivial_poisson,eq:non_dim_trivial_forcelaw} become
\begin{gather}
  \label{eq:non_dim_poisson_small_pots}
  \tilde{\chargedens} = - \tilde{\bn}^2\tilde{\elpot},
  \\
  \label{eq:non_dim_forcelaw_small_pots}
  \num{0} = \tilde{\elpot} + \dielrelop \tilde{\chargedens},
\end{gather}
where $\dielrelop$ is defined as the dielectric operator
\begin{equation}
  \label{eq:dielectric_operator}
  \dielrelop
  = 1- \sum_{n=0}^1 \tilde{\varGamma}^{2n}_{+-}\cdot \tilde{\bn}^{2n}
  =
  \num{1}
  - \frac{\Vint}{\Eth}
  - \frac{2}{\pi} 
  \frac{\Vint}{\Eth}
  \frac{\Eel}{\Eth} 
  \tilde{\bn}^2.
\end{equation}
In the absence of molecular repulsion, $\Vint=\num{0}$, the dielectric
operator reduces to the canonical, scalar-valued dielectric parameter
$\dielrelop\to \num{1}$.

Furthermore, quantities similar to $\dielrelop$ also arise in the
liquid state theory of classical statistical mechanics. This expansion corresponds to a small wave vector expansion of the dielectric function expressed as a correlation function of the molecular dipole densities (see,  e.g., Ref.\citenum{hansen2006theory}).
		
In the following sections, we show that the gradient expansion,
\cref{eq:non_dim_forcelaw_small_pots,eq:non_dim_poisson_small_pots},
allows significant insights into the competing effects of the
interactions $\Vint,\Eel,\Eth$, and predicts the EDL structure as
a function of the temperature, dielectricity, ion-size, and
interaction strength.
	
\section{Mean Steric Effect: Charge Saturation}
\label{sec:underscreening}
	
\begin{figure*}[!htb]
  \includegraphics[width=1.8\columnwidth]{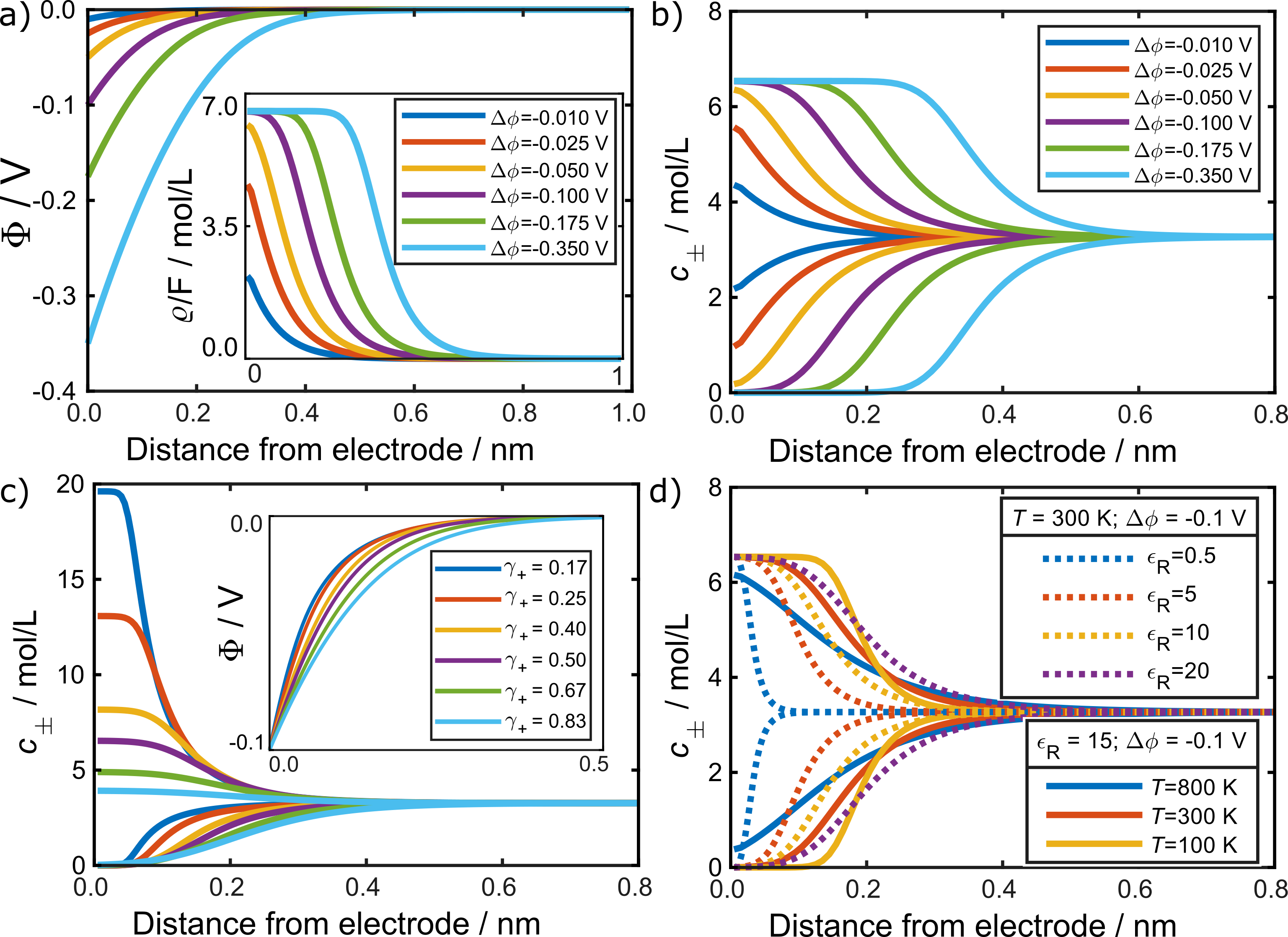}
  \caption{Simulation results of the EDL structure perpendicular to
    the electrode-electrolyte interface for a binary IL for $\Vint=\num{0}$ (see
    \cref{eq:non_dim_trivial_poisson,eq:dyn_tt_gradexpansion}
    ). If
    not mentioned otherwise, $\tempr=\SI{300}{\kelvin}$,
    $\dielrel=\num{15}$, and $\varDelta\phi=\SI{-0.1}{\volt}$. a)
    Profiles of electric potential and charge density (inset) for
    different electrode potentials $\varDelta\phi$. b) Concentration
    profiles of the anions and cations for different electrode
    potentials $\varDelta\phi$. c) Concentration profiles for
    different volume ratios $\gammaup_+$. The inset shows the
    corresponding electric potential. d) Concentration profiles for
    varying dielectric constants (dashed lines) and temperatures
    (solid lines).}
  \label{fig:crowding_numeric_all}
\end{figure*}
	
In this section, we neglect non-local interactions, $\Vint=0$, and
discuss the EDL structure of the electrolyte due to bulk effects alone
based on $F^{\ce{b}}$ (see \cref{eq:interaction_extension}). In this
way, we reveal the competition between Coulombic ordering and entropic
disordering, i.e., diffusion.
	
Toward this aim, we consider the system of equations constituted by the
Poisson \cref{eq:non_dim_trivial_poisson} and
\cref{eq:non_dim_trivial_forcelaw} subject to $\Vint=0$,
\begin{equation}
  \label{eq:bulk_cont_trnsport_eq}
  0
  =
  \tilde{\elpot}
  - \gammaup_+
  \ln\left[
    \tilde{\conc}_-/\tilde{\conc}^{\ce{b}}
  \right]
  \\
  + \gammaup_-
  \ln\left[\tilde{\conc}_+/\tilde{\conc}^{\ce{b}}
  \right].
\end{equation}
	
First, in \cref{sec:numer-invest}, we solve this system of equations
numericaly. We supplement this investigation by an analytic analysis,
and focus on the two limiting regimes of large and small electric
potentials. In \cref{sec:small-potentials-bulk} we discuss the case
$\tilde{\elpot}\ll \num{1}$, and in \cref{sec:therm-limit-boltzrt}, we
discuss the case $\tilde{\elpot}\gg \num{1}$. For the special case of
symmetric ion-species ($\gammaup_\pm = 0.5$), we derive analytic
solutions for the electric field
$\tilde{\boldsymbol{E}}(\tilde{\elpot})$, and for the charge density
$\tilde{\chargedens}(\tilde{\elpot})$ as functions of the electrolyte
electric potential in the SI (see \cref*{sec:SImean-volume-effect}).
	
\subsection{Simulations}
\label{sec:numer-invest}
	
\Cref{fig:crowding_numeric_all} shows numeric results for the system
of
\cref{eq:non_dim_trivial_poisson,eq:bulk_cont_trnsport_eq}. \Cref{fig:crowding_numeric_all}a
and \cref{fig:crowding_numeric_all}b illustrate screening profiles of
the electric po\-ten\-tial, the charge density, and the ion
con\-cen\-tra\-tions for varying electrode potentials $\varDelta\phi$.
	
Apparently, the application of a negative electrode potential
($\varDelta\phi<\num{0}$) polarizes the electrolyte. The electric
potential (see \cref{fig:crowding_numeric_all}a) is continuous across
the interface and decays smoothly towards the electroneutral bulk
region. The inset of \cref{fig:crowding_numeric_all}a shows that, for
low electrode potentials, the charge density decays exponentially. A
similar behavior can be observed in \cref{fig:crowding_numeric_all}b
for the concentrations. The concentration of positive counter-ions
increases toward the interface, whereas negative ions get
depleted. Apparently, the electrolyte screens the electrode potential
by accumulation of counter-ions. However, above
$\varDelta\phi\approx \SI{-0.05}{\volt}$, the counter-ion
concentration saturates near the interface. A further increase of
$\varDelta\phi$ broadens the EDL.
	
This behavior can be explained by the mean volume effect. The
application of a negative potential $\varDelta\phi$ implies that
positive ions accumulate near the interface, and negative ions
deplete. However, the Euler equation for the volume,
\cref*{eq:SI_elyte_eq_of_state}, implies the saturation concentration
$\conc_+^{\ce{sat}} = \num{1}/\pmv_+$. Once the accumulated species
reaches this saturation, the screening mechanism transitions from
increasing the concentration at the interface to broadening the width
of the EDL. The simulated EDL approaches a thickness of
\SI{0.6}{\nano\meter} at $\varDelta\phi\approx \SI{-0.05}{\volt}$,
which is significantly wider than predicted by the canonical
Debye-H\"uckel-theory with the Debye length
$L_{\ce{D}}\approx \SI{0.7}{\angstrom}$ (see
\cref{eq:def_debye_length}). This phenomenon is typically denoted
``crowding''.\cite{bazant}
	
Because the saturation concentration depends upon the molar volume,
$\conc_\alpha^{\ce{sat}}=\num{1}/\pmv_\alpha$, the partial molar
volumes directly affect the screening
behavior. \Cref{fig:crowding_numeric_all}c shows numerical results for
the ionic concentrations at different volume ratios
$\gammaup_\pm=\pmv_\pm/\pmv$ (in which $\pmv$ is kept fixed). The EDL is
thinner for smaller $\gammaup_+$, as this allows for tighter packing
of cations.
	
The effect of temperature $\tempr$ and dielectricity $\dielrel$ on the
EDL structure is illustrated in \cref{fig:crowding_numeric_all}d. The
screening is more effective for smaller values of $\dielrel$ and the
EDL width increases with increasing magnitude of $\dielrel$. This is
in qualitative agreement with the screening behavior for dilute
solutions as predicted by the Debye-H\"uckel theory. Likewise, the EDL
becomes more diffuse with increasing temperature because of the
disordering effect of thermal motion. The observed effects of $\tempr$
and $\dielrel$ highlight the competing interplay between the
electrostatic effect of charge ordering and the disordering effect of
entropy.
	
To summarize, the simulations show two distinct regimes of EDL
structure. First, for large electrode potentials,
$\Delta\tilde{\phi}\gg \num{1}$, the charge is saturated near the
interface. Second, near the electroneutral bulk region, where
$\Delta\tilde{\phi}\ll \num{1}$, the charge density decays
exponentially. These two distinct EDL structures, charge saturation
and ex\-po\-nen\-tial de\-crease, correspond to two disjointed
electrolyte regimes, $\lvert\tilde{\elpot}\rvert\gg\num{1}$ and
$\lvert\tilde{\elpot}\rvert\ll\num{1}$, respectively.
	
\subsection{Asymptotic Analysis}	
\label{sec:story-telling}

\begin{figure}[!htb]
  \includegraphics[width=0.9\columnwidth]{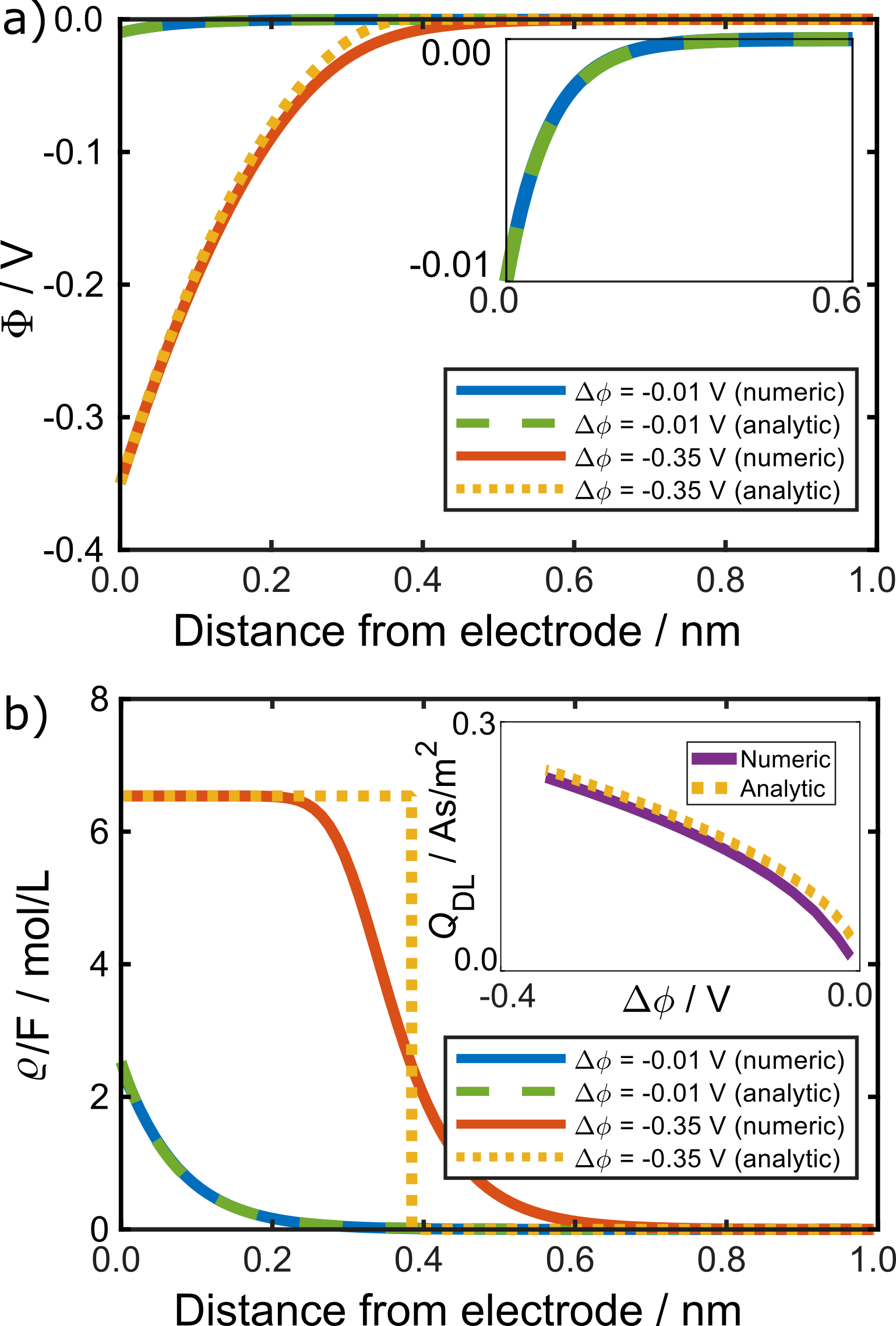}
  \caption{Comparison of the asymptotic analysis of the EDL structure
    with numerical results obtained from \cref{eq:non_dim_trivial_poisson,eq:dyn_tt_gradexpansion}
    ($\tempr=\SI{300}{\kelvin}$, $\Vint=\num{0}$ and $\dielrel=\num{15}$). We consider
    two different values of the interface potential, corresponding to
    the regimes of small and large potentials at
    $\varDelta\phi=\SI{-0.01}{\volt}$
    ($\Delta\tilde{\phi}=\num{-0.4}$), and
    $\varDelta\phi=\SI{-0.35}{\volt}$
    ($\Delta\tilde{\phi}=\num{-13.5}$). a) Profiles of the electric
    potentials as predicted by the analytical approximation (see
    \cref{eq:small_phi_bulk_sol_phi,eq:phi_ttozero}).  The inset
    highlights the region close to the interface for the case
    $\varDelta\phi=\SI{-0.01}{\volt}$. b) Profiles of the charge
    density as predicted by the asymptotic analysis in
    \cref{eq:small_rho_bulk_sol_phi,eq:rho_t_to_zero}. The inset
    compares numerical and analytical values for the total charge
    in the EDL (see \cref*{sec:SI_asympt-analys-large} in the SI).}
  \label{fig:crowding_numeric_analytix_1}
\end{figure}
	
The simulation results in \cref{sec:numer-invest} motivate our
procedure for analysing the EDL structure. First, we study the EDL far
away from the electrode close to the electroneutral bulk (large
$\tilde{\xcomp}$). For this purpose, we expand the stationary equations
around the electroneutral bulk for small charge densities,
$\tilde{\chargedens}\ll\num{1}$. According to
\cref{eq:bulk_cont_trnsport_eq}, this regime corresponds to small
dimensionless potentials $\lvert\tilde{\elpot}\rvert\ll\num{1}$. Note
that this coincides with the high-temperatures regime because 
$\tilde{\elpot} = \elpot\cdot Fz_+/R\tempr$ (see
\cref{sec:small-potentials-bulk}).
	
Second, we analyze the behavior close to the electrode (small
$\tilde{\xcomp}$) at large electrode potentials
$\lvert\Delta\tilde{\phi}\rvert\gg \num{1}$, where the electrolyte
potential satisfies $\lvert\tilde{\Phi}\rvert\gg\num{1}$. This
corresponds to low temperatures (see
\cref{sec:therm-limit-boltzrt}).
	
\subsubsection{Asymptotic Analysis: Small Potentials
  \texorpdfstring{$\lvert\tilde{\elpot}\rvert\ll\num{1}$}{LG}}
\label{sec:small-potentials-bulk}
	
As outlined above, we begin our analytic examination of the EDL in the
limit of small dimensionless potentials
$\lvert\tilde{\elpot}\rvert\ll\num{1}$. Our idea is to approach the
EDL from the electroneutral bulk region along the direction of
decreasing $\tilde{\xcomp}$. Thus, we use the expansion of
ionic concentrations around the bulk electrolyte in
\cref{eq:log_taylor},
$\tilde{\conc}_{\pm} = \tilde{\conc}^{\ce{b}}\pm
\tilde{\chargedens}\gammaup_{\mp} \approx \tilde{\conc}^{\ce{b}}$.
	
For this aim, we insert $\tilde{\chargedens}$ from
\cref{eq:non_dim_poisson_small_pots} into
\cref{eq:non_dim_forcelaw_small_pots} for $\dielrelop=\num{1}$,
yielding
\begin{equation}
  \label{eq:bulk__eq_small_phi}
  \tilde{\bn}^2 \tilde{\elpot}
  = \tilde{\elpot}.
\end{equation}
With the boundary conditions discussed above,
$\tilde{\elpot}(\num{0}) = \varDelta\tilde{\phi} =
Fz_+\varDelta\phi/R\tempr$ and
$\tilde{\elpot}(\tilde{\xcomp}\to\infty) = \num{0}$, we obtain the
solution
\begin{gather}
  \label{eq:small_phi_bulk_sol_phi}
  \tilde{\elpot}
  = \varDelta\tilde{\phi}\cdot \exp(-\tilde{\xcomp}),\\
  \label{eq:small_rho_bulk_sol_phi}
  \tilde{\chargedens} = - \varDelta\tilde{\phi}\cdot
  \exp(-\tilde{\xcomp}).
\end{gather}
The corresponding dimensionalized electrolyte potential,
\begin{equation}
  \label{eq:small_phi_bulk_sol_phi_1}
  \elpot
  = \varDelta\phi\cdot \exp(-\xcomp/L_{\ce{D}}),
\end{equation}
decays exponentially with damping parameter $1/L_{\ce{D}}$. Thus, the
decay length in the limit $\lvert\tilde{\elpot}\rvert\ll \num{1}$ is
the Debye length $L_{\ce{D}}$ (see \cref{eq:def_debye_length}).
	
In \cref{fig:crowding_numeric_analytix_1}, we compare the analytic
predictions for this limit (dashed green lines) with the numeric
results (blue lines) for different electrode potentials. Apparently,
the analytical and numerical results for the electric potential and the
charge density are in excellent agreement for small electrode
potentials, $\Delta\tilde{\phi}\approx \num{-0.4}$, when the condition
$\lvert\tilde{\elpot}\rvert\ll\num{1}$ is fulfilled everywhere.

In the SI, we derive the expressions for the total
EDL-surface-charge-density and the differential capacitance (see
\cref*{sec:SI_asympt-analys-small}).

\subsubsection{Asymptotic Analysis: Large Potentials
  \texorpdfstring{$\lvert\tilde{\elpot}\rvert\gg\num{1}$}{LA}}
\label{sec:therm-limit-boltzrt}
	
Next, we discuss the EDL in the limit of large potentials
$\lvert\tilde{\phi}\rvert\gg\num{1}$. This regime can be found for
large electrode potentials $\Delta\tilde{\elpot}\gg \num{1}$ close to
the electrode/electrolyte interface. As
$\tilde{\elpot}=Fz_+\varDelta\phi/R\tempr$, this analysis is exact in
the limit of zero temperature, $\tempr=\num{0}$.
	
In this case, the logarithmic terms in \cref{eq:bulk_cont_trnsport_eq}
must compensate the diverging potential term $\tilde{\elpot}$. Because of
the mean volume constraint (see \cref*{eq:SI_elyte_eq_of_state}), one of
the logarithmic terms is diverging if one species is depleted and the
other species saturates,
$\chargedens(\xcomp=\num{0}) = F z_{\satindex} /
\pmv_{\satindex}$. Here, we denote the saturating species by the index
$\satindex$. Because electric potentials are continuous across
interfaces, the saturation species $\satindex$ is uniquely determined
by the sign of the electrode potential,
$\ce{sign}(z_{\satindex})=-\ce{sign}\,(\varDelta\tilde{\phi})$.
	
Therefore,
$\tilde{\chargedens}^{\ce{sat}} = -
\ce{sign}(\varDelta\tilde{\phi})\cdot \tilde{\conc}^{\ce{sat}}$ solves
\cref{eq:bulk_cont_trnsport_eq}, where
$\tilde{\conc}^{\ce{sat}}=
\tilde{\conc}^{\ce{b}}/\gammaup_{\satindex}$. Upon integration of the
Poisson \cref{eq:non_dim_poisson_small_pots}, we find
\begin{equation}
  \label{eq:phi_ttozero}
  \tilde{\elpot}(\tilde{\xcomp}) = \varDelta\tilde{\phi}\cdot 
  \left(\num{1}-\tilde{\xcomp}/\tilde{L}_{\ce{EDL}}\right)^2,
\end{equation}
where the width of the EDL depends on the electrode po\-ten\-tial,
$\tilde{L}_{\ce{EDL}} = \sqrt{ \num{2}
  \gammaup_{\satindex}\lvert\varDelta\tilde{\phi} \rvert
  /\tilde{\conc}^{\ce{b}}}$. Thus, the dimensionalized EDL length is
\begin{equation}
  \label{eq:L_DL_ttozero}
  L_{\ce{EDL}} = L_{\ce{D}} \tilde{L}_{\ce{EDL}}
  =  \sqrt{ \num{2}\ionsize^3 \lvert \varDelta \tilde{\phi} \rvert
    \gammaup_{\satindex} 
    k_{\ce{B}} T\diel\dielrel/(ez_+)^2} .
\end{equation}
Apparently, the decay length increases with ion size because of the
mean-volume effect. Also, it scales with the asymmetry
$\sqrt{\gammaup_{\satindex}}$; \textit{i.e.}, it is small for small
screening species.  Comparison with the Debye screening-length shows
that $L_{\ce{EDL}}>L_{\ce{D}}$ in the limit of small temperatures
$\tempr$ or large potentials
$\lvert\varDelta\tilde{\phi}\rvert\gg\num{1}$.
	
In the limit of vanishing temperature, $\tempr=\num{0}$, the charge profile
is box shaped and is determined by the screening-length
$L_{\ce{EDL}}$,
\begin{equation}
  \label{eq:rho_t_to_zero}
  \tilde{\chargedens} 
  =  -\theta(\tilde{L}_{\ce{EDL}} -
  \tilde{\xcomp}) \cdot \ce{sign}(\varDelta\tilde{\phi})
  \cdot \tilde{\conc}^{\ce{sat}},
\end{equation}
where $\theta$ is the Heaviside function. In the SI, we calculate the
analytic expressions for the total charge in the EDL and for the
differential capacitance in this limit (see
\cref*{sec:SI_asympt-analys-large}).
	
In \cref{fig:crowding_numeric_analytix_1}, we compare the analytical
predictions for this limit (dashed yellow lines) with the numerical
results (solid red lines) for $\Delta\tilde{\phi}=\num{-13.5}$. The
box-shaped charge-profile is in good qualitative agreement with the
numerical results because it almost predicts the correct width of the
EDL. However, the transition from saturation to the bulk-state is more
diffuse in the numerical profile. This is due to the entropic, thermal
influence, which ``washes out'' the box. Nevertheless, the inset shows
that the charge in the EDL, as predicted by the analytical
approximation, is quantitatively in good agreement with the numerical
results. We note that this profile and its temperature dependence are
reminiscent of the Fermi distribution.
	
\section{Non-Local Interactions: Charge Oscillations}
\label{sec:inter-symm-ions}
	
In this section, we discuss the influence of non-local interactions
($\Vint\ne 0$) on the structure of the EDL. As in
\cref{sec:underscreening}, we discuss the two limiting cases of small
and large potentials separately.
	
\subsection{Static Asymptotic Analysis}
\label{sec:math-analys}
	
\subsubsection{Asymptotic Analysis: Large Potentials
  \texorpdfstring{$\lvert\tilde{\elpot}\rvert\gg\num{1}$}{LB}}
\label{sec:interactions_large-potentials}
	
Let us first discuss the regime of diverging electrolyte potentials
$\lvert\tilde{\elpot}\rvert\to\infty$. In this limit, the interaction
contribution cannot compensate the diverging electrolyte potential in
\cref{eq:non_dim_trivial_forcelaw}. The logarithmic terms are
diverging if one species is depleted, and we recover the same results as
described in \cref{sec:therm-limit-boltzrt} for the case of vanishing
interaction contributions $\Vint=0$.
	
\subsubsection{Asymptotic Analysis: Small Potentials
  \texorpdfstring{$\lvert\tilde{\elpot}\rvert\ll\num{1}$}{LC}}
\label{sec:interactions_small-potentials}
	
\begin{figure*}[!htb]
  \includegraphics[width=1.929\columnwidth]{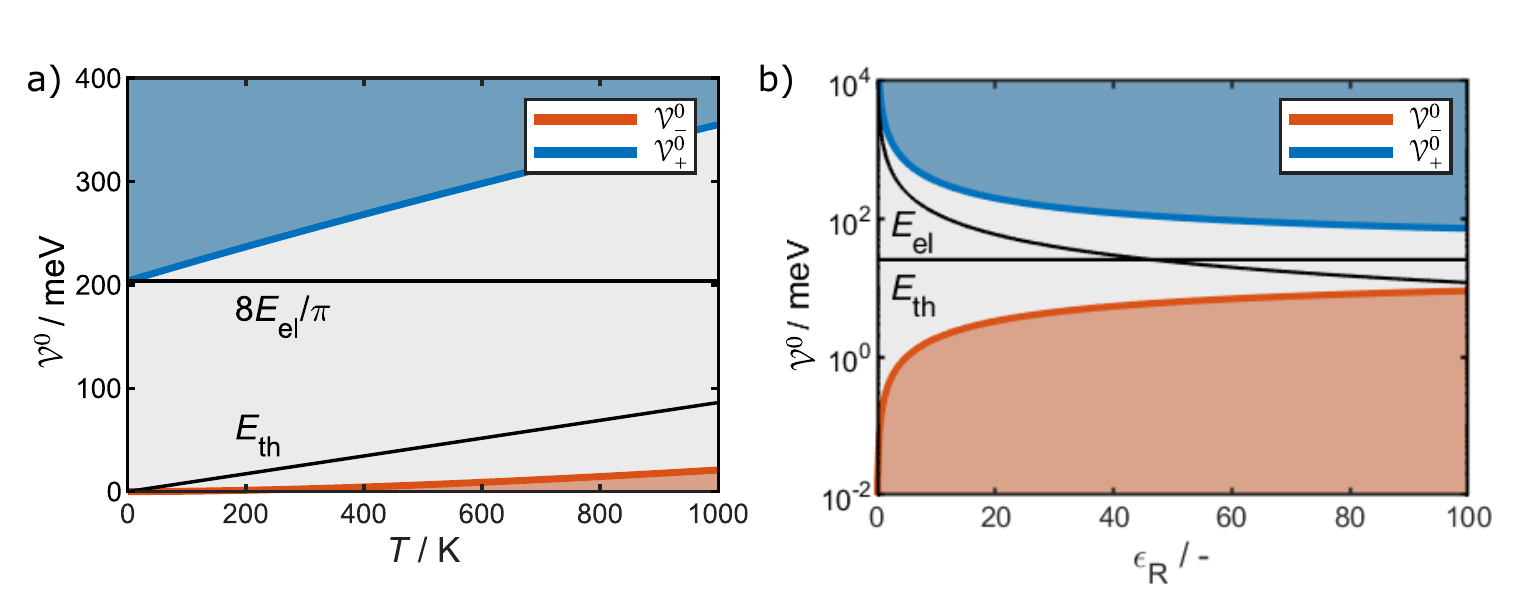}
  \caption{Phase spaces for EDL structure as functions of $T$ and
    $\dielrel$ for equally sized ions ($\gammaup_\pm = \num{0.5}$),
    see \cref{eq:critstrengths}. a) Critical interaction energies
    $\Vint_\pm$ as function of temperature (here,
    $\dielrel = \num{15}$, $\ionsize = \SI{1.3}{\nano\meter}$). b)
    Critical interaction energies $\Vint_\pm$ as function of
    dielectricity (here, $T=\SI{300}{\kelvin}$ and
    $\ionsize=\SI{1.3}{\nano\meter}$). Three phases are present:
    exponentially damped charge density (shaded red), decaying
    oscillatory charge density (shaded gray), quasi-crystalline
    (shaded blue).}
  \label{fig:phasespaces_various}
\end{figure*}
	
In this section, we consider the full theory with molecular repulsion
in the regime of small potentials
$\lvert\tilde{\elpot}\rvert\ll \num{1}$.  As outlined above, we expand
the interaction free energy in gradients of the charge density, and
restrict our analysis to the first two perturbation modes $n=\num{0}$ and
$n=\num{1}$. With the assumption of small charge densities
$\tilde{\chargedens}\ll 1$, we derived linear equations above
(\cref{eq:non_dim_poisson_small_pots,eq:non_dim_forcelaw_small_pots}),
which we rephrase in matrix form,
\begin{gather}
  \label{eq:nondim_fundamental_system}
  \left(
    \begin{matrix}
      \tilde{\bn}^2\tilde{\chargedens}
      \\
      \tilde{\bn}^2\tilde{\elpot}
    \end{matrix}
  \right) = \tilde{\boldsymbol{\mathscr{A}}\ } \cdot \left(
    \begin{matrix}
      \tilde{\chargedens}
      \\
      \tilde{\elpot}
    \end{matrix}
  \right), \intertext{where}
  \label{eq:nondim_matrix_A}
  \tilde{\boldsymbol{\mathscr{A}}\ } = \left(
    \begin{matrix}
      \displaystyle \frac{\pi\Eth^2}{\num{2}\Eel\Vint} \left(\num{1} -
        \frac{\Vint}{\Eth} \right) & \quad \quad \quad & \displaystyle
      \frac{\pi\Eth^2}{\num{2}\Eel\Vint}
      \\
      \\
      - \num{1} & & 0
    \end{matrix}
  \right).
\end{gather}
We solve \cref{eq:nondim_fundamental_system} via the eigenvalue
decomposition with the eigenvalues $\tilde{\upalpha}_{1,2}$ of the
matrix in \cref{eq:nondim_matrix_A}. These are determined by the relative
magnitudes of the three competing energy scales, $\Eth$, $\Eel$, and
$\Vint$,
\begin{equation}
  \label{eq:eigenvaluepm}
  \tilde{\upalpha}_{1,2}
  =
  - \frac{\pi}{\num{4}}
  \cdot \frac{\Eth}{\Eel}
  \cdot \left[
    \num{1} - \frac{\Eth}{\Vint}
    \mp
    \sqrt{
      \left(1- \frac{\Eth}{\Vint}\right)^2
      - 
      \frac{\num{8}}{\pi}
      \cdot \frac{\Eel}{\Vint}
    }
  \right] .
\end{equation}
Each eigenvalue $\tilde{\upalpha}_i$ gives rise to a dimensionless
wave-vector,
\begin{equation}
  \label{eq:k_from_aplha}
  \tilde{k}_{1,2} = \sqrt{\tilde{\upalpha}_{1,2}}.
\end{equation}
These determine the general solution of \cref{eq:nondim_fundamental_system} together with the eigenvectors 
$\tilde{\boldsymbol{a}}^{\tilde{\upalpha}_{i}} =
(\tilde{a}^{\tilde{\upalpha}_{i}}_{1},
\tilde{a}^{\tilde{\upalpha}_{i}}_{2})^T =
(-\tilde{\upalpha}_{i},1)^T$,
\begin{equation}
  \label{eq:general_solution}
  \left(
    \begin{matrix}
      \tilde{\chargedens}
      \\
      \tilde{\elpot}
    \end{matrix}
  \right)
  =
  \left(
    \begin{matrix}
      \boldsymbol{a}^{\tilde{\upalpha}_{1}} &
      \boldsymbol{a}^{\tilde{\upalpha}_{2}}
    \end{matrix}
    \vphantom{  \left(
        \begin{matrix}
          A_1 e^{\tilde{k}_1\tilde{\xcomp}} + A_2
          e^{-\tilde{k}_1\tilde{\xcomp}}
          \\
          A_3 e^{\tilde{k}_2\tilde{\xcomp}} + A_4
          e^{-\tilde{k}_2\tilde{\xcomp}}
        \end{matrix}
      \right)}
  \right)
  \cdot
  \left(
    \begin{matrix}
      A_1 e^{\tilde{k}_1\tilde{\xcomp}} + A_2
      e^{-\tilde{k}_1\tilde{\xcomp}}
      \\
      A_3 e^{\tilde{k}_2\tilde{\xcomp}} + A_4
      e^{-\tilde{k}_2\tilde{\xcomp}}
    \end{matrix}
  \right).
\end{equation}
The expansion coefficients $A_i$ are determined by boundary conditions
and physical arguments. Apparently, the corresponding wave-vectors are
functions
$\tilde{k}_{1,2}(\tempr,z_\alpha,\dielrel,\pmv_\alpha,\mathscr{F}_{\alpha\beta})$,
which determine the structure of the EDL,
\begin{equation*}
  \label{eq:profile_cases}
  \tilde{k}_{1,2} \in
  \begin{cases}
    \mathbb{R}, & \textnormal{exponential damping},
    \\
    \mathbb{R}+ \mathrm{i} \cdot \mathbb{R}, & \textnormal{damped
      oscillations},
    \\
    \mathrm{i} \cdot \mathbb{R}, & \textnormal{oscillations}.
  \end{cases}
\end{equation*}
Thus, the EDL structure depends upon the relative magnitudes of the
energies $\Eth$, $\Eel$, and $\Vint$ via \cref{eq:eigenvaluepm}. In
particular, the classification of $\tilde{k}_\pm$ depends upon the sign of the
root
\begin{equation}
  \label{eq:Wroot}
  \Wroot
  = \left(\num{1} - \Eth/\Vint\right)^2
  - \num{8}\Eel/\Vint\pi
\end{equation}
appearing in \cref{eq:eigenvaluepm}. Thus, the critical values
$\Vint_\pm$, defined by the condition $\Wroot(\Vint_\pm)=0$, determine
the thresholds for the transition between the phases of EDL structure,
\begin{equation}
  \label{eq:critstrengths}
  \Vint_\pm
  =
  \Eth + \num{4}\Eel/\pi
  \pm \num{2}\sqrt{\num{2}\Eel(\num{2}\Eel/\pi+\Eth)/\pi}.
\end{equation}
Thus, with \cref{eq:critstrengths} we can draw the phase diagram for the 
EDL structures. Since $\num{0}<\Vint_-<\Vint_+$, there are three
phases. In the SI, we discuss each case in great detail (see \cref*{sec:SIphase-analysis}). Next, we give
a short description of each phase.
	
\textit{Phase 1:} $\num{0}\leq \Vint\leq\Vint_-$. In this regime,
$\tilde{\upalpha}_{1,2}\geq \num{0}$, which implies a real-valued
wave-vector. Thus, this phase corresponds to exponentially damped
profiles,
$\tilde{\chargedens}\propto
\exp(-\tilde{k}_{\mathbb{R}}\tilde{\xcomp})$. A harmonic analysis of
the root $\sqrt{\Wroot}$ reveals that (see \cref*{sec:SIlimiting-cases})
\begin{equation}
  \label{eq:k_limit_v_to_zero}
  \lim_{\Vint\to \num{0}}\tilde{k}_{1}
  =\infty
  \quad \text{and}
  \quad
  \lim_{\Vint\to 0}\tilde{k}_{2}
  =\num{1}.
\end{equation}
Thus, solutions with the damping parameter $\tilde{k}_1$ vanish
quickly and are rendered as unphysical, whereas the limit of vanishing
interactions for $\tilde{k}_2$ reproduces the ``bulk''-expansion for
$\tilde{\elpot}\ll 1$ from \cref{sec:small-potentials-bulk}, see
\cref{eq:small_phi_bulk_sol_phi}.
	
\textit{Phase 2:} $\Vint_-<\Vint<\Vint_+$. In this regime, $\Wroot<0$,
and, thus, the root \cref{eq:Wroot} becomes complex. Therefore, the
wave-vector has non-vanishing real and imaginary parts,
$\tilde{k}_{1,2} \in \mathbb{R}\times \mathrm{i} \cdot
\mathbb{R}$. This corresponds to charge-profiles of exponentially
damped oscillations,
\begin{equation}
  \label{eq:damped_oscillations}
  \tilde{\chargedens} \propto
  \exp(-\tilde{k}_{\mathbb{R}}\tilde{\xcomp})\cdot
  \cos(\tilde{k}_{\mathbb{C}}\tilde{\xcomp}).
\end{equation}
	
\begin{figure}[!htb]
  \includegraphics[width=\columnwidth]{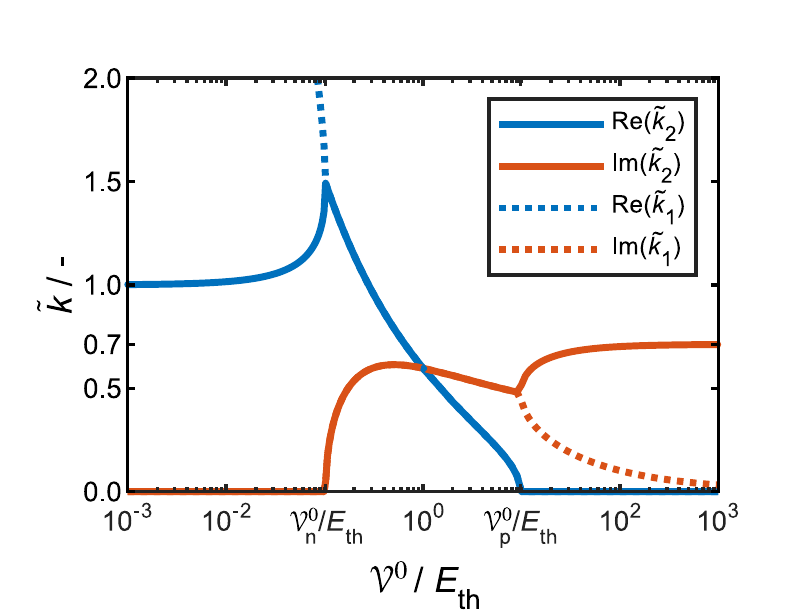}
  \caption{Real and imaginary parts of the non-dimensionalized
    wave-vector $\tilde k=k\cdot L_{\ce{D}}$ as function of the
    relative magnitude of the energy scales $\Vint$ and $\Eth$ (see
    \cref{eq:eigenvaluepm,eq:k_from_aplha}). Here,
    $T=\SI{300}{\kelvin}$, $\dielrel=\num{15}$, and
    $\ionsize=\SI{1.3}{\nano\meter}$.}
  \label{fig:phasespaces_wavevector}
\end{figure}
	
\textit{Phase 3:} $\Vint_+\leq\Vint$. In this regime, both eigenvalues
are real but negative, $\tilde{\upalpha}_{1,2}\leq0$. Therefore,
$k_\pm\in \mathrm{i}\cdot \mathbb{R}$, which corresponds to undamped
oscillatory profiles. The limiting case for indefinitely strong
interactions yields
\begin{equation}
  \label{eq:terminal_k}
  \lim_{\Vint\to\infty} k_{1,2}
  = \lim_{\Vint\to\infty} \tilde{k}_{1,2} / L_{\ce{D}}
  = \begin{cases}
    0,\\
    \pm \mathrm{i}  \cdot 2\pi/\ionsize.
  \end{cases}
\end{equation}
Thus, the result for $\lim_{\Vint\to\infty} k_{2}$ reproduces the
experimental findings obtained by AFM-measurements that the
wave-lengths $\uplambda\approx \ionsize/\num{2}\pi$ of the observed
oscillations scale with the size of molecules
$\ionsize$.\cite{C7CP08243F,li2013effect} Apparently, the
incompressibility of ions in our model prevents a further decrease of
the wavelength.
	
Thus, the critical values $\Vint_\pm$ constitute exactly the
boundaries between the different EDL phases.
	
In \cref{fig:phasespaces_various}, we illustrate the phase space of
EDL structures as functions of temperature and dielectricity (in
\cref*{sec:phase-space-scre} we also show the phase space as function
of ion-size, ion-asymmetry and valency). Apparently, three distinct
phases of EDL structures are present. The exponentially damped EDL phase
corresponds to the regions below $\Vint_-$ (red line), whereas the
damped-oscillatory EDL phase corresponds to the regions between the
blue and red lines. Finally, the undamped oscillatory EDL phase
corresponds to the regions above $\Vint_+$ (blue
line). \Cref{fig:phasespaces_various}a illustrates that temperature
$\tempr$ and hardcore-interactions $\Vint$ are in competition and that the
critical interaction strengths $\Vint_\pm$ increase with increasing
temperature $\tempr$, i.e., thermal energy $E_\text{th}$.
	
\Cref{fig:phasespaces_various}b reveals the influence of
dielectricity, i.e., electro-static forces, on the EDL
phases. Apparently, the damped oscillatory phase becomes narrower for
ILs with larger dielectricity $\dielrel$, i.e., smaller electric
energy $E_\text{el}$.
	
We note that the exponentially damped regime for small interaction
strengths (small compared to electrostatic and thermal energy)
corresponds to the EDL structure found in
\cref{sec:small-potentials-bulk} in the absence of
hardcore-interactions. However, as can be inferred from
\cref{fig:phasespaces_various}a, this phase is hardly present for
reasonable temperatures.
	
\Cref{fig:phasespaces_wavevector} shows the non-dimensionalized
wave-vector as function of the relative energy scale $\Vint/\Eth$
(where $T{=}\SI{300}{\kelvin}$, $\dielrel{=}\num{15}$, and
$\ionsize{=}\SI{1.3}{\nano\meter}$). For small interaction energies
$\Vint<\Vint_-$ the wave-vector is real,
$\tilde{k}=\tilde{k}_{\mathbb{R}}$, which corresponds to exponentially
damped profiles.  In particular, the static profile at $\Vint\to 0$
reproduces the case described in \cref{sec:story-telling} where the
exponential profile is determined by the scales $\Eth$ and $\Eel$
alone, \text{i.e.} $\tilde{k}=1$. Apparently, $\tilde{k}_{\mathbb{R}}$
increases with $\Vint$ up to the threshold $\Vint_-$, beyond which
which it starts to decrease. Thus, the EDL has minimal extension at
$\Vint=\Vint_-$. This suggests that the increasing strength of the repulsive
ion-correlations compresses the screening-layer. Once the hardcore
potential exceeds $\Vint_-$, the system overscreens, \textit{i.e.}, the
ion-layers begin to oscillate. The damping parameter
$\text{Re}(\tilde{k})$ vanishes exactly when $\Vint=\Vint_+$,
\textit{i.e.}, when the system transitions into 
nano-segregation of the ion species. Interestingly, the frequency of
the oscillations $\text{Im}(\tilde{k})$ exhibits a local maximum and
minimum in the regime of damped oscillations. Furthermore, 
$\text{Im}(\tilde{k})$ attains its maximal value in the limit of
prevailing interaction strength $\Vint\to\infty$.
	
In the SI (see
\cref*{sec:SItriv-order-expans,sec:SIline-order-expans}) we
investigate the influence of the individual perturbation modes
$\varGamma^0_{12}$ and $\varGamma^2_{12}$ on the phase diagram. As it
turns out, neglecting all but the zeroth-order correction
$\varGamma^0_{12}$ results in a binary phase diagram comprising only
exponentially damped profiles and undamped, oscillatory profiles. In
contrast, neglecting the zeroth-order correction, and taking only the
first non-trivial order $\varGamma^2_{12}$ into account, results in a
binary phase-diagram comprising only exponentially damped profiles and
damped oscillatory profiles. This is the case for MFTs based on the
BSK-framework. Thus, for the ``complete'' set of the three different
phases, both perturbation modes $\varGamma^0_{12}$ and
$\varGamma^2_{12}$ are neccessary.

Interestingly, for the pathologic case of negative
in\-ter\-act\-ion\--strengths $\Vint<0$, the phase space reduces to the two
screening phases of exponentially damped profiles and undamped
oscillatory profiles. This follows straightforwardly from
\cref{eq:eigenvaluepm} (see also the discussions in
\cref*{sec:SItriv-order-expans,sec:SIline-order-expans}).

\subsection{Dynamic Asymp\-to\-tic Analysis: Linear
    Stability Analysis}
\label{sec:linstabanalysis}

In this section, we complement the static analysis of
\cref{sec:math-analys} by an analytic analysis of the dynamic
transport equation in the gradient description
(\cref{eq:dyn_tt_gradexpansion}).

For this purpose, we perform a linear stability analysis and consider the limit of small potentials, $\lvert\tilde{\elpot}\rvert\ll 1$. Thus, the logarithmic terms can approximated as in
\cref{eq:log_taylor}, and \cref{eq:dyn_tt_gradexpansion} becomes
\begin{equation}
\label{eq:timedependent_tt}
\p_{\tilde{t}}\tilde{\chargedens}
= -\tilde{\bn}^2 \left(
\tilde{\elpot} + \dielrelop \tilde{\chargedens} \right).
\end{equation}
We expand the electric potential around an uniform bulk-state $\tilde{\elpot}^{\ce{b}}$,
\begin{equation}
  \label{eq:lin_stab_exp}
  \tilde{\elpot} = \tilde{\elpot}^{\ce{b}} + \sum_{i=1}^\infty
  \epsilon^i \cdot \tilde{\elpot}_i.
\end{equation}
Here, the equilibrium state is determined by the electroneutral
bulk-condition $\tilde{\elpot}^{\ce{b}}=0$ and $\tilde{\chargedens}^{\ce{b}}=0$. Thus, the first order
perturbation takes the form
\begin{equation}
\label{eq:lin_stab_firstorder_exp}
\tilde{\elpot}_1 
= \exp\left[\tilde{s}\tilde{t}\right]\cdot\cos\left[\tilde{k}\tilde{\xcomp}\right].
\end{equation}
Here, the wave-number $\tilde{k}$ determines the spatial distribution of the dimensionless perturbation $\tilde{\epsilon}^1\ll 1$, and the
parameter $\tilde{s}$ measures the temporal growth rate of this
perturbation.

We restrict our analysis to probing the linear stability and
substitute \cref{eq:lin_stab_exp} and the Poisson equation into
\cref{eq:timedependent_tt}. Next, we collect  terms up to the first order in
the perturbation mode $\tilde{\epsilon}^1$, which yields a dispersion
relation for the growth rate of the perturbation,
\begin{equation}
  \label{eq:dispersion_relation}
  \tilde{s}(\tilde{k})
  = -1 - \left(
    1 -  \frac{\Vint}{\Eth} 
    + \frac{2}{\pi} \frac{\Vint}{\Eth} \frac{\Eel}{\Eth}( \tilde{k})^2
  \right) \cdot (i\tilde{k})^2.
\end{equation}

The uniform state is stable under perturbations if and only if
$\tilde{s}<0$. This defines an instability onset $\tilde{k}^{\ce{c}}$
for the wave-numbers
\begin{equation}
  \label{eq:stability_onset}
  \tilde{k}^{\ce{c}}_{1,2} 
  = \pm
  \frac{1}{2} \sqrt{\pi(\Vint\Eth- \Eth^2)/\Eel\Vint}.
\end{equation}
The corresponding stability criterion
$\tilde{s}(\tilde{k}^{\ce{c}}_{1,2}){<}0$ determines the phase
boundary at which the bulk of the IL-electrolyte becomes
unstable. This 
stability threshold exactly equals the phase boundary between the damped oscillatory
phase and the nano-segregated phase (see \cref{eq:critstrengths}), 
\begin{equation}
  \label{eq:crit_stability_crit_en}
  \Vint_{\ce{+}}
  = \Eth+4\Eel/\pi
  + 2 \sqrt{2\Eel(2\Eel/\pi+\Eth)/\pi}.
\end{equation}

Thus, for interaction energies $\Vint{>}\Vint_{+}$ the bulk state of
the system becomes unstable and phase separation emerges. The initial
cause for the structure-formation can be driven by external agents, or
boundary conditions, e.g., by the application of an electric
potential to an IL/electrode interface.

This stability analysis complements the static analysis, and
rationalizes the emergence of phase separation into ionic layers
occurring at interaction energies above $\Vint_+$.

\subsection{Validation With Simulation}
\label{sec:numeric-validation}

Our goal in this section is to compare the results of our asymptotic
analysis (see
\cref{eq:eigenvaluepm,eq:k_from_aplha,eq:general_solution} in
\cref{sec:interactions_small-potentials}), with numeric simulations of
the completely-coupled system subject to the two theoretical
descriptions (integral description
\cref{eq:non_dim_trivial_poisson,eq:dyn_tt_integral}, and gradient
description
\cref{eq:non_dim_trivial_poisson,eq:dyn_tt_gradexpansion}).

\begin{figure}[!tb]
\includegraphics[width=\columnwidth]{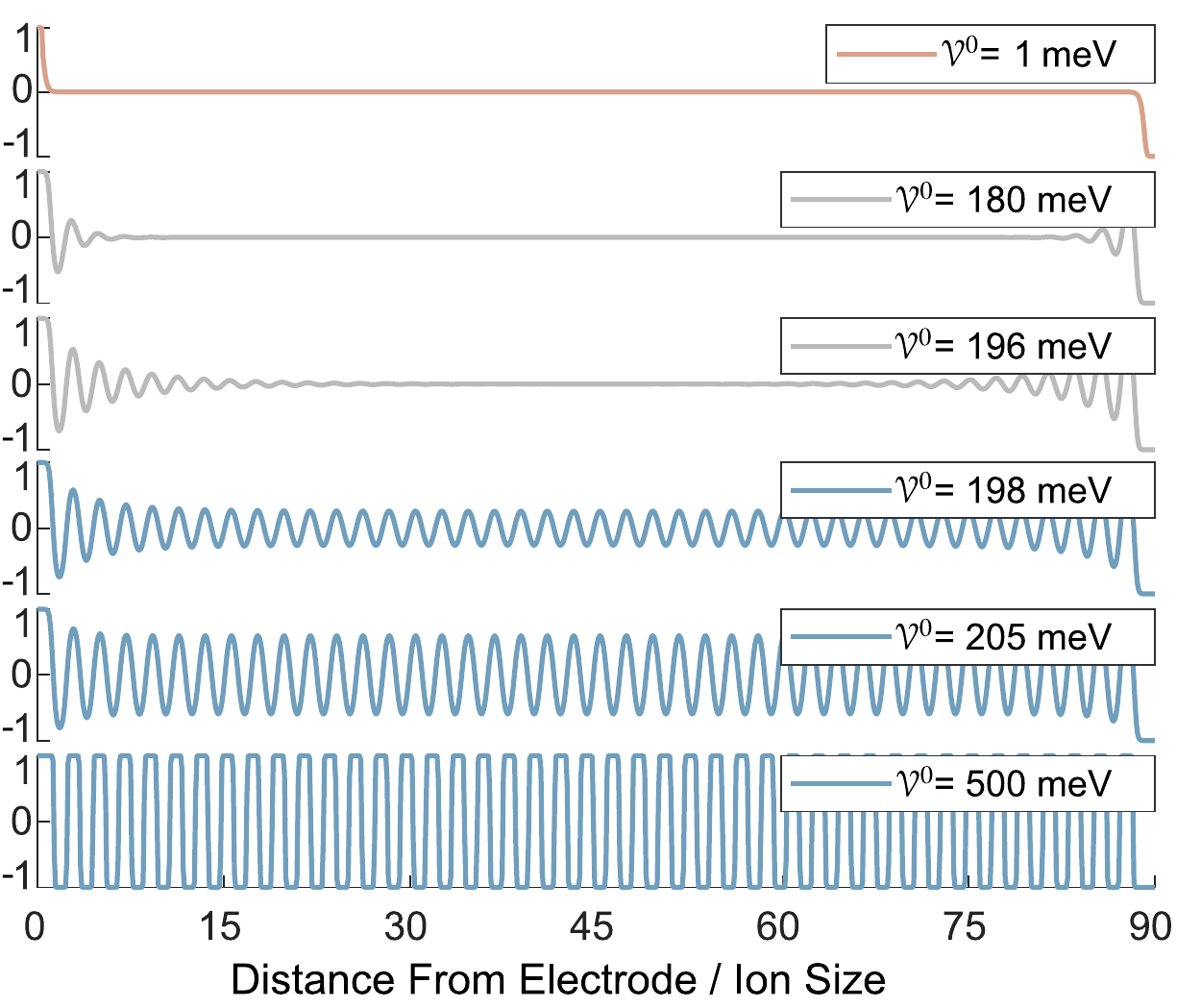}
\caption{Screening profiles of the charge density $\tilde{\chargedens}$, as obtained from numerical simulations of the integral description (see \cref{eq:non_dim_trivial_poisson,eq:dyn_tt_integral}) for different values $\Vint$. The $y$-axis is scaled from
  \num{-1} to \num{1}, where $\lvert\tilde{\chargedens}\rvert=1$
  corresponds to charge saturation.}
\label{fig:integral_ode_phse_overview}
\end{figure}

\begin{figure}[!tb]
\includegraphics[width=\columnwidth]{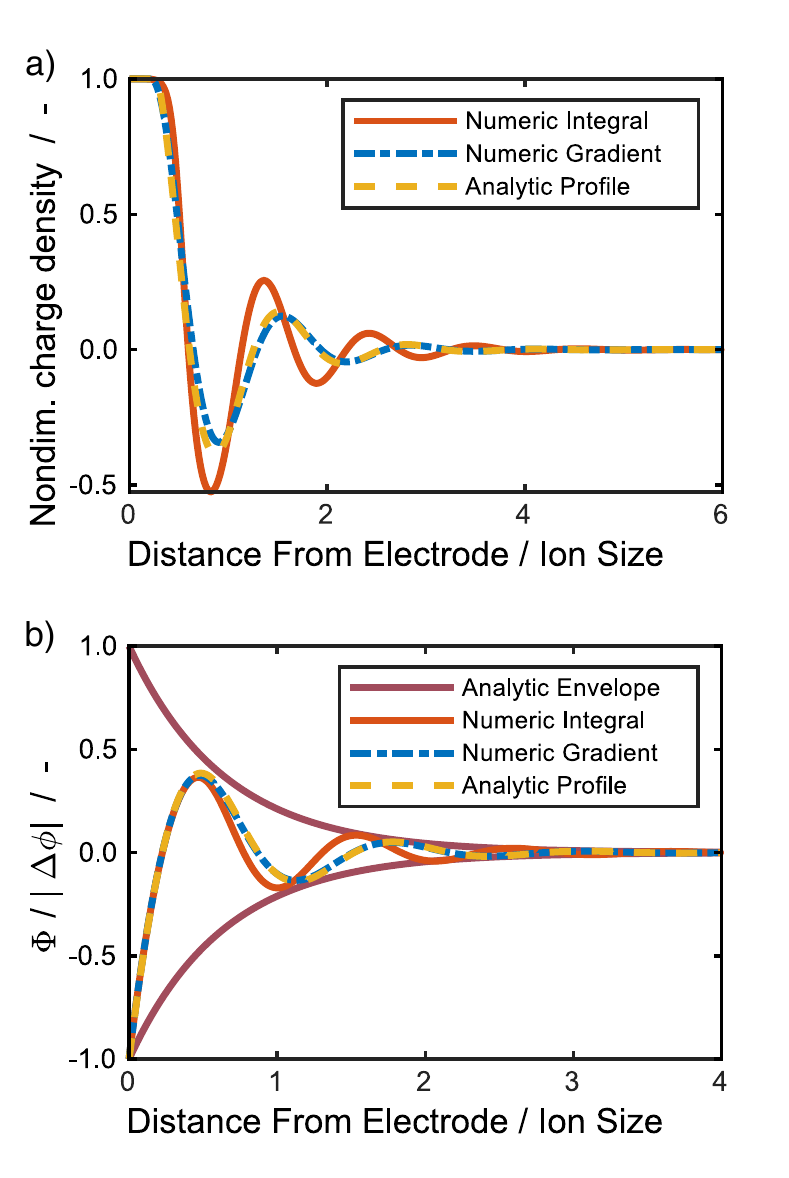}
\caption{Results for electric potential and charge density as obtained from numerical simulations of the integral description (\cref{eq:non_dim_trivial_poisson,eq:dyn_tt_integral}), and of the gradient description  (\cref{eq:non_dim_trivial_poisson,eq:dyn_tt_gradexpansion}), and as predicted by the analytic predictions (see \cref{eq:eigenvaluepm,eq:k_from_aplha,eq:general_solution}) at $\Vint=\SI{180}{\milli\electronvolt}$.}
\label{fig:combi_intermediate_phase}
\end{figure}

\begin{figure}[!tb]
\includegraphics[width=\columnwidth]{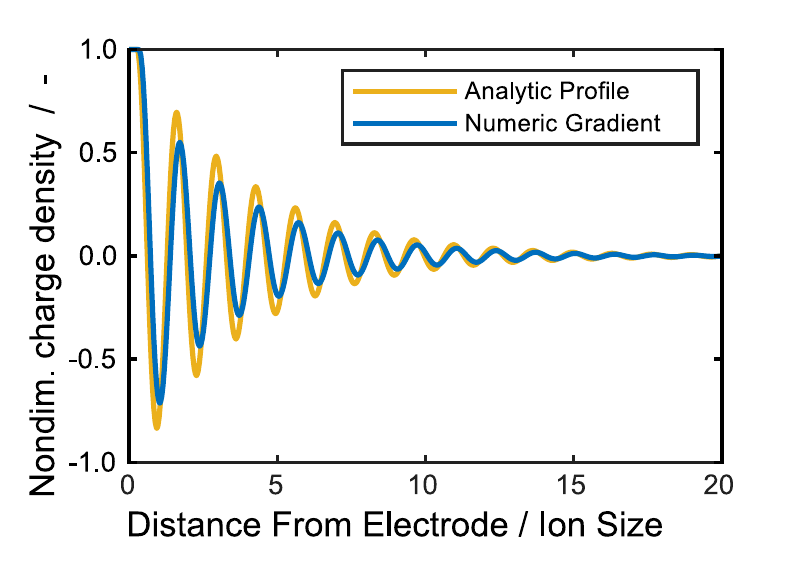}
\caption{Screening profile for the charge density $\tilde{\chargedens}$ obtained from numerical simulations with
  respect to the gradient description
  (\cref{eq:non_dim_trivial_poisson,eq:dyn_tt_gradexpansion}), and
  according to the analytic description (see
  \cref{eq:eigenvaluepm,eq:k_from_aplha,eq:general_solution}), at
  $\Vint=\SI{250}{\milli\electronvolt}$.}
\label{fig:combi_intermediate_prtrbtn_nexttocrystallinephase}
\end{figure}

\begin{figure}[!htb]
\includegraphics[width=\columnwidth]{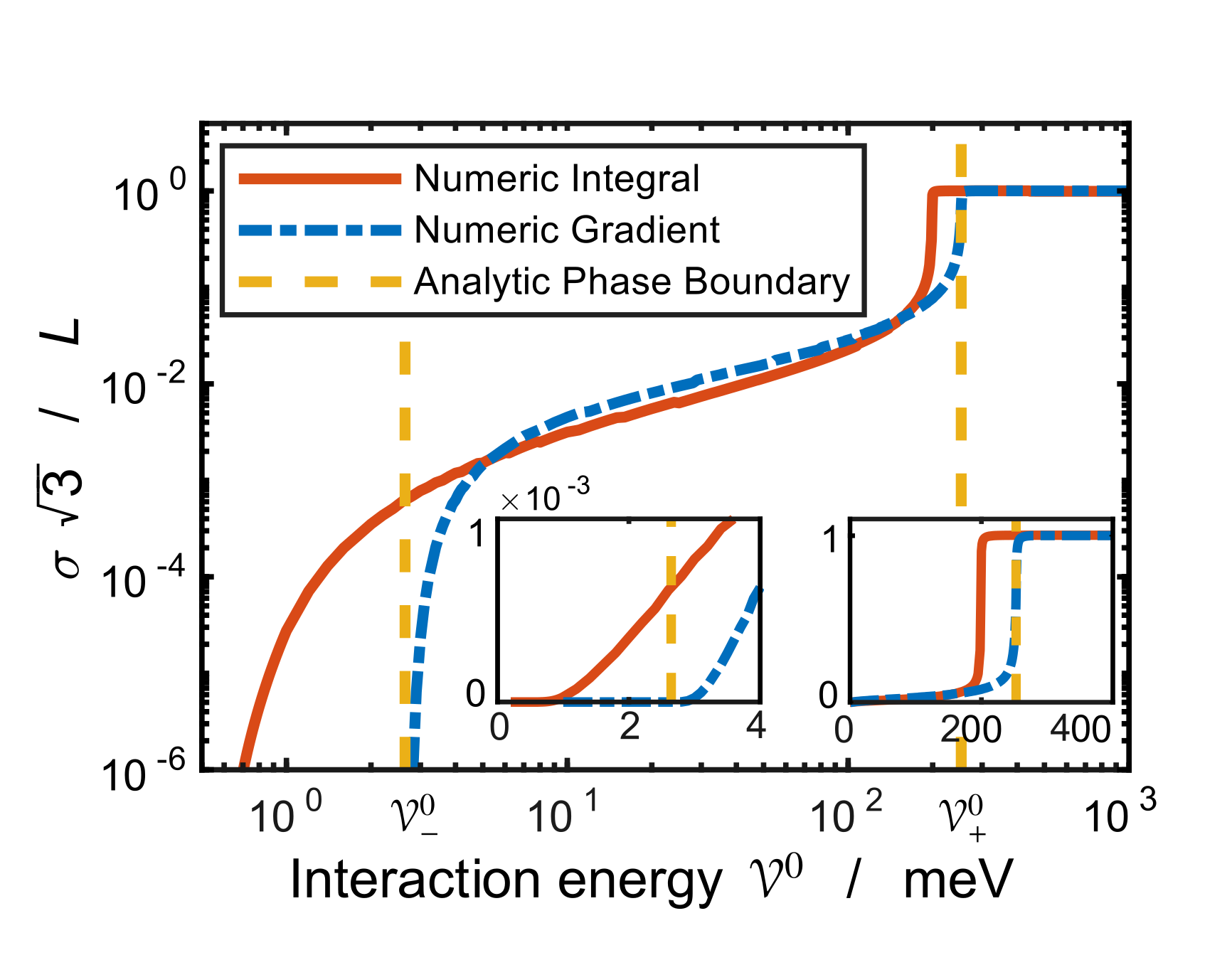}
\caption{Meta analysis of the interfacial profiles for some thousand simulations. The dashed vertical yellow lines show the phase boundaries $\Vint_\pm$ (see \cref{eq:critstrengths}). The dashed and solid red / blue lines show the peak-variance of the complete set of simulations as defined by \cref{eq:variance}, with respect to the integral description (see \cref{eq:non_dim_trivial_poisson,eq:dyn_tt_integral}), and with respect to the gradient-description (see \cref{eq:non_dim_trivial_poisson,eq:dyn_tt_gradexpansion}). The left inset shows the onset of the oscillations at small interaction energies. The right inset shows the variance in a non-logarithmic setting, which highlights the occurrence of phase-transitions.}
\label{fig:overview_multisimulations}
\end{figure}

We start our numerical investigations with an overview of the screening profiles for the charge density at different values $\Vint$, as obtained from the integral description \cref{eq:non_dim_trivial_poisson,eq:dyn_tt_integral} (see \cref{fig:integral_ode_phse_overview}). Next, we compare our different EDL-descriptions in detail for two different energies $\Vint$, see \cref{fig:combi_intermediate_phase,fig:combi_intermediate_prtrbtn_nexttocrystallinephase}. Finally, we generalize these exemplary findings via a systematic study
over the complete phase-space of interaction energies, see \cref{fig:overview_multisimulations}. This provides a clear illustration of the complete set of 
phase-transitions which the system undergoes, and highlights the consistency between the three descriptions.

All simulations were performed for a symmetric cell set-up, where the
IL-electrolyte is located within two oppositely charged, blocking
interfaces separated by a distance of $L_{\ce{cell}}=\SI{60}{\nano\meter}$. The electrode on the left side is negatively charged with
$\varDelta\phi = \SI{-100}{\milli\volt}$, whereas, on the right side, the electrode is positively charged with $\varDelta\phi = \SI{100}{\milli\volt}$. Since charge saturation begins roughly at 
$\varDelta\phi=\SI{70}{\milli\volt}$, see
\cref{fig:crowding_numeric_all,fig:crowding_numeric_analytix_1}, the 
charge distribution can safely be assumed saturated adjacent to the
interfaces (\textit{i.e.} $\lvert\tilde{\chargedens}\rvert=1$). The electrolyte is
considered to consist of symmetric ions ($\gammaup_\pm=\num{.5}$)
of size $\ionsize=\SI{1.3}{\nano\meter}$. Hence, the
cell-geometry allows a ``maximal'' number of roughly \num{90}
ions. In addition, we assume room temperature,
$T=\SI{300}{\kelvin}$, and $\dielrel=\num{15}$. The phase boundaries corresponding to these parameters as predicted by our analytic description are
$\Vint_-=\SI{3}{\milli\electronvolt}$ and
$\Vint_+=\SI{253}{\milli\electronvolt}$ (see \cref{eq:critstrengths}).

\Cref{fig:integral_ode_phse_overview} shows the numerical results of the
charge density for the integral description
(\cref{eq:non_dim_trivial_poisson,eq:dyn_tt_integral}), where $\Vint$
takes values across two orders of magnitude. First, at  
$\Vint=\SI{1}{\milli\electronvolt}$ the profile shows charge saturation near the two electrified electrodes, $\tilde{\chargedens}(\xcomp{=}0)=1$ and $\tilde{\chargedens}(\xcomp{=}L_{\ce{cell}})=-1$. Near both electrodes, the profile  decays exponentially towards the electroneutral bulk ($\tilde{\chargedens}=0$). This corresponds to the profiles which we
discussed in great detail in \cref{sec:numer-invest}. Because
$\Vint<\Vint_-$, this is in accordance with the analytic
prediction. The next two profiles show results for interaction
energies within the intermediate phase, $\Vint_-<\Vint<\Vint_+$. Both simulations show damped oscillatory profiles, where the 
long-ranged oscillatory profiles span many
nanometers. Apparently, the oscillations in the profile for
$\Vint=\SI{196}{\milli\electronvolt}$ extend almost across the entire
cell. A slight increase by
$\SI{2}{\milli\electronvolt}$ to $\Vint=\SI{198}{\milli\electronvolt}$
causes the profile to transition into a crystalline phase with undamped oscillatory shape. Note that the amplitudes between the electrodes are smaller than unity, \textit{i.e.} the bulk region consists of mixed ion layers with one  dominant ion species. An increase to
$\Vint=\SI{204}{\milli\electronvolt}$ enhances the amplitudes of the oscillations further, \textit{i.e.} enhances segregation of ion-species. The last plot shows the corresponding profile for a significantly enhanced interaction energy ($\Vint=\SI{500}{\milli\electronvolt}$). Here, the amplitudes of the oscillations have reached saturation ($\lvert\tilde{\chargedens}\vert=1$), and the electrolyte has
transitioned into a crystalline phase consisting of alternating pure  ion layers. In \cref*{fig:detail_crystalline_phase} (see \cref*{sec:SI-crystalline-limit}), we highlight that the ionic layers coincide exactly with the ion size $\ionsize$. Thus, with increasing energies $\Vint$, the interfacial structure increases into the bulk electrolyte, until the bulk itself gets nanostrucured by the layering of the ion species. This phase transition occurs rapidly within a few $\SI{}{\milli\electronvolt}$.

Apparently, the numerical results for the integral description confirm the existence of three different screening phases. However, quantitative deviations between our descriptions are present. As we show in the SI (see \cref*{fig:SI-peak-numbers}) the phase transitions from exponential decay to damped oscillations occurs roughly at $\Vint=\SI{2}{\milli\electronvolt}$. In addition, as can be inferred from \cref{fig:integral_ode_phse_overview}, the transition from damped oscillations to undamped oscillations appears at  $\Vint=\SI{200}{\milli\electronvolt}$. Hence, both phase boundaries are slightly shifted to smaller values, compared to the analytical predictions $\Vint_-=\SI{3}{\milli\electronvolt}$ and  $\Vint_+=\SI{253}{\milli\electronvolt}$ (see \cref{eq:critstrengths}). Thus, the analytical prediction, which is based on the gradient description, slightly underestimates the influence of $\Vint$, when compared with $\Eth$ and $\Eel$. This can be attributed to the fact that the gradient description is an approximation based on only the first two perturbation modes, whereas the integral description comprises all modes.

Next, we give a quantitative comparison between the numerical results of the two theoretical descriptions, and the analytical predictions. Here, we restrict our discussion to the interaction energy
$\Vint=\SI{180}{\milli\electronvolt}$, \textit{i.e.} the intermediate
phase of damped oscillations. \Cref{fig:combi_intermediate_phase}
shows the profiles for the charge density and electrolyte electric potential as obtained from the numeric simulations, and as predicted by the analytic description for the first few nanometers of the left half-cell. \Cref{fig:combi_intermediate_phase}a illustrates the charge distribution adjacent to the negatively charged electrode. The dashed blue line shows the screening profile obtained from the gradient description, which exhibits a damped oscillatory shape. This confirms the analytic prediction for this interaction energy. The dashed yellow line shows the resulting analytic profile. 
Note that the analytic prediction in
\cref{sec:inter-symm-ions} does not capture charge-saturation, but only determines the damping-parameter and the  oscillation-frequency of the
screening-profile. However, in \cref{sec:therm-limit-boltzrt}, we derived an analytic prediction for the saturation-width $L_{\ce{EDL}}$, valid close to the interface  (see \cref{eq:L_DL_ttozero}). Hence, in order to reconstruct the "complete" profile, we supplement the contribution emerging from the bulk  \cref{eq:general_solution}, valid far away from the interface, by constant charge-saturation $\tilde{\chargedens}=1$ spanning over the width $L_{\ce{EDL}}$. Apparently, the analytic and numeric results of the gradient description are quantitatively in very good agreement. Finally, the solid red line in \cref{fig:combi_intermediate_phase}a shows the numerical results for the integral description. In accordance with the results shown in \cref{fig:integral_ode_phse_overview}, these results reproduce the analytically predicted screening phase, but the oscillations are more pronounced. Hence, the influence of the interaction energy $\Vint$ is more dominant in the integral description than in the gradient description. Next, in \cref{fig:combi_intermediate_phase}b) we show the profiles for the normalized electrolyte electric potential. The dashed blue line shows the profile due to the gradient description. It is in accordance with the charge-profile shown in \cref{fig:combi_intermediate_phase}a, see also \cref{eq:non_dim_poisson_small_pots,eq:log_taylor}. Again, we reconstruct the analytic profile by supplementing the profile \cref{eq:phi_ttozero}, valid close to the interface, by the profile \cref{eq:general_solution}, valid towards the electroneutral bulk. Apparently, the analytic results are quantitatively in very good agreement with the results stemming from the gradient description. The red line shows the profile as obtained from the integral description. Like in \cref{fig:combi_intermediate_phase}a for the charge density, the oscillations are slightly enhanced when compared with the gradient description. In addition, the brown solid lines show the analytic envelopes for the screening. Apparently, it captures both numerical results qualitatively very well. Interestingly, the differences in electrolyte potential between the three descriptions depend on electrode potential near the electrode as shown in \cref*{fig:figure_SI_phidependence_overscreening} (see \cref*{sec:SI_phi_dependence_overscreening}). However, the qualitative agreement is independent from the boundary conditions.

In \cref{fig:combi_intermediate_prtrbtn_nexttocrystallinephase}, we show results for the charge distribution at  enhanced interaction energy $\Vint=\SI{250}{\milli\electronvolt}$, \textit{i.e.}, close to the phase boundary  $\Vint_+=\SI{253}{\milli\electronvolt}$. As can be inferred from \cref{fig:integral_ode_phse_overview}, the integral description has already transitioned to the phase of undamped oscillations for this interaction energy. Hence, we only show the screening profile  as obtained from numerical simulation of the gradient description (solid blue line), and compare it with the analytic prediction (yellow line). Apparently, in accordance with the analytic prediction, the numerical  profile has a damped oscillatory shape, where the oscillations extend over roughly \num{40} ion sizes. This highlights the influence of the enhanced interaction energy, see also  \cref{fig:combi_intermediate_phase}. Overall, the analytic profile shown here is in nice agreement with the numeric results. 

Finally, we conduct a quantitative comparison between the two EDL-discriptions and the analytic description across multiple orders of magnitude of $\Vint$. To address this goal, we examine simulation results of roughly 4000 EDL-simulations across the parameter range
from $\SI{0.1}{\milli\electronvolt}$ up to $\SI{500}{\milli\electronvolt}$. As above, we apply $\Delta\phi=\SI{\pm100}{\meV}$ at the electrodes such that we can safely assume charge saturation near the interfaces. 

We evaluate the simulation results by extracting two characteristic properties. First, we count the number of peaks $\tn_{\ce{peaks}}(\Vint)$ appearing in each screening profile. Due to charge saturation, a minimal number of two peaks always occurs. At most, roughly 90 ion-layers fit into the cell geometry of length $L_{\ce{cell}}=\SI{60}{\nano\meter}$. We present the number of peaks occurring in the full cell as function of the interaction strengths in the SI (see \cref*{fig:SI-peak-numbers} in \cref*{sec:simulation-results}).

However, beyond the number of  peaks, we want to evaluate the peak amplitudes, too. For this purpose, we investigate the  peak variance  $\sigma(\Vint)$ of the left half-cell, defined by
\begin{equation}
  \label{eq:variance}
  \sigma^2= 
    \frac{\sum_{i=1}^{\tn_{\ce{peaks}}}\tilde{\chargedens}_i\cdot
    \left(\xcomp_i\right)^2}{\sum_{j}^{\tn_{\ce{peaks}}}\tilde{\chargedens}_j}.
\end{equation}
Here, $\xcomp_i$ is the discrete location of the $i$-th peak
$\tilde{\chargedens}_i=\lvert\tilde{\chargedens}(\tilde{\xcomp}_i)\rvert$
(such that $0\leq\tilde{\chargedens}_i\leq 1$, where $\tilde{\chargedens}_i=1$ corresponds to a saturated peak, \textit{i.e.} a pure ion-layer) appearing in the
profile of the charge density. In the SI (see \cref*{sec:SI_meta_anlss_peak_nmbrs}), we show analytically that $\sigma$ converges to $L_{\ce{cell}}/\sqrt{3}$ if the set of simulation-energies comprises energies $\Vint\gg\Vint_+$. For such interaction energies, the bulk electrolyte has
transitioned into a crystalline phase composed of nano-segregated
ion-layers (see  \cref{fig:integral_ode_phse_overview}). 

\Cref{fig:overview_multisimulations} shows the results for variance
$\sigma$ normalized to its maximum $L_{\ce{cell}}/\sqrt{3}$ in
logarithmic scale. In this figure, the vertical dashed yellow lines
indicate the phase-boundaries $\Vint_\pm$, as predicted by the
analytic description (see \cref{eq:critstrengths}). The left inset
shows the simulation-results for small values $\Vint$, and the right
inset comprises the overall results in a non-logarithmic
representation highlighting the transition.  The blue dashed line
shows the results for $\sigma$ according to the gradient description
(see \cref{eq:non_dim_trivial_poisson,eq:dyn_tt_gradexpansion}). At
small interaction energies $\Vint<\Vint_-$, the variance is zero. This
corresponds to an exponentially damped, non-oscillatory screening
profiles (note that the only peak, due to charge saturation, is
located at $\tilde{\xcomp}_i=0$). The variance starts increasing
exactly at $\Vint_-$ (see also the left inset). This corresponds to an
increasing number of damped oscillations, where the amplitudes of the
peaks also increase with $\Vint$.  Finally, at $\Vint_+$, the variance
converges to it's constant limiting value $L_{\ce{cell}}/\sqrt{3}$
(see also the right inset). In this energy-regime, the bulk
electrolyte consists completely of pure ion-layers.  Altogether, these
results reproduce exactly the phases as predicted by the analytic
description.

The red curve shows the results for $\sigma$ according to the integral description (see \cref{eq:non_dim_trivial_poisson,eq:dyn_tt_integral}). In contrast to the gradient description, the variance starts increasing from zero at roughly $\Vint=\SI{1}{\meV}$, \textit{i.e.} before the analytically predicted phase boundary $\Vint_-$ (see also the left inset). Hence, the phase transition from exponentially damped screening-profiles to damped oscillatory screening-profiles is slightly shifted to smaller energies. Next, the variance increases exponentially up to roughly $\Vint=\SI{200}{\meV}$, above which it transitions into the constant limiting value $L_{\ce{cell}}/3$. Altogether, the
phase boundaries of the integral description still exhibit a qualitatively good agreement with the analytic predictions, although being slightly shifted to smaller values. Apparently, this behaviour is due to the cumulative effect of the integral-term in \cref{eq:dyn_tt_integral}, which comprises all interaction modes. In contrast, we consider only the first two modes (n=0 and n=1) of the gradient expansion in \cref{eq:dyn_tt_gradexpansion}.

\section{Multi-Scale Methodology}

In this section, we highlight the relation of our model to theories on
smaller and larger length scales. We discuss in
\cref{sec:relat-md-meth}, based on basic concepts from liquid state
theory, how atomistic simulations can directly parametrize our
theory. Next, in \cref{sec:from-non-equilibrium} we sketch the
phenomenologic BSK continuum approach for the description of ILs near
electrified interfaces and illustrate its relation to our work. In addition, we state the relation of our framework to AFM-experiments in the \cref*{sec:relat-exper-meth} ( see also Ref.\citenum{C7CP08243F}).

\subsection{From Molecular Dynamics to Non-\-Equi\-li\-brium Thermodynamics}
\label{sec:relat-md-meth}

Here, we explain how the parameters of our continuum theory can be rigorously
calculated with quantum chemistry, i.e., DFT and MD.

Ab-initio DFT calculations predict the forces between ions and
molecules by calculating their electronic structure. The DFT-generated
force-fields are the focal quantity for MD simulations, \cite{refId0}
which calculate the classical trajectories of ions and
molecules. Results from MD simulations are often interpreted via
profiles of the radial distribution function $g(r)$.

Liquid state
theory,\cite{hansen2006theory} connects this atomistic description to
thermodynamic concepts and scattering
experiments.\cite{doi:10.1080/00268977900102331} On the one hand, the
radial distribution function allow a straightforward comparison with
the structure factor $S$ from scattering
experiments.\cite{doi:10.1098/rsta.2018.0413,PhysRevE.97.012610} On
the other hand, the density distribution function $g(r)$ can be used to calculate different correlation functions. By subtracting its asymptotic value
follows the so-called total correlation function used in integral
equation theories (IETs), $h(r)=g(r)-1$.\cite{henderson2009attractive}
In IETs, the pairwise total correlation function $h$ relates to the
the direct correlation function $c^{(2)}$, used in classical
density functional theory (cDFT), via the Ornstein-Zernike
relation,\cite{doi:10.1063/1.4993175}
\begin{equation}
  \label{eq:ornstein_zernike}
  h(r) = c^{(2)}
  + \massdens_{\ce{b}} \int \ce{d}r^\prime  c^{(2)}(\lvert
  r-r^\prime\rvert ) \cdot h(r^\prime).
\end{equation}
In cDFT, the direct pair correlation functions
$ c^{(2)}_{\alpha\beta}$ account for pairwise interactions between
two ions of species $\alpha$ and $\beta$, \textit{i.e.}, the excess
free energy due to pairwise
ion-interactions.\cite{https://doi.org/10.1002/adts.201900049} Thus,
they can be obtained via the twofold functional derivative of
$\Fint$,\cite{doi:10.1080/00268977900102331} \textit{i.e.}, via our
interaction potential $\mathscr{F}$ (see
\cref{eq:rel_intpot_corrltn_fct}),
\begin{equation}
   c^{(2)}_{\alpha\beta}(\lvert
  r-r^\prime\rvert )
  = - \frac{1
  }{
    k_{\ce{B}}T(N_{\ce{A}})^2} \cdot 
  \mathscr{F}_{\alpha\beta}(\lvert r-r^\prime\rvert).
\end{equation}

To summarize, DFT determines force fields for MD, MD determines $g(r)$
for liquid state theory, $g(r)$ determines $ c^{(2)}$ via the
Ornstein-Zernike relation, $c^{(2)}$ determines $\Fint$ and
generates our non-equilibrium thermodynamic theory.

The dynamic properties of our theory can be determined from atomistic
simulations, too. These dynamic properties are encoded in the Onsager
coefficients,\cite{schammer2020theory} which can be measured
experimentally.\cite{doi:10.1021/acs.macromol.0c02545} The Onsager
coefficients can be determined by MD simulations (``Green Kubo
relations'').\cite{https://doi.org/10.1002/aic.17091,gotze1989generalised,latz_diss}

\subsection{From Non-Equilibrium Thermodynamics to Phenomenologic BSK
  Theory}
\label{sec:from-non-equilibrium}
	
Now, we compare our thermodynamically consistent continuum approach
with the phenomenologic theory proposed by Bazant, Storey, and
Kornyshev (BSK), a seminal MFT-approach for ILs near electrified
interfaces.\cite{bazant} In their continuum model of the EDL, BSK
incorporate ion-correlations using a modified linear dielectric
relation
$ \bar{\boldsymbol{D}}=\hat{\bar{\varepsilon}} \boldsymbol{E}$ between
the electrostatic fields $\bar{\boldsymbol{D}}$ and $\boldsymbol{E}$,
where $\hat{\bar{\varepsilon}} = \dielrel\diel (1-\ell_{\ce{c}}\bn^2)$
is their dielectric operator. The second order gradient term in
$\bar{\chargedens}$ accounts for non-local ion-interactions, being
effectively short-ranged with correlation length $\ell_{\ce{c}}$. This
Ansatz yields a modified Poisson equation,
$\hat{\bar{\varepsilon}}\bn^2{\elpot} = -\bar{\chargedens}$. The
chemical potential connects electric potential and charge density. Finally,
\begin{equation}
    \hat{\bar{\varepsilon}}\bn^2{\elpot} = \elpot.
\end{equation}
holds in the limit of small potentials $\tilde\elpot$.

Our model conceptually differs from BSK theory. Since we incorporate
electrostatic correlations in the free energy, non-local
ion-interactions enter the set of equations via the chemical
potentials. This implies that the MFT-quantities appearing in the BSK
description, $\bar{\boldsymbol{D}}$ and
$\bar{\chargedens}=\bn\bar{\boldsymbol{D}}$, differ from the
corresponding quantities $\chargedens$ and $\boldsymbol{D}$ appearing
in our formalism. In contrast to the "mean field charge density"
$\bar{\chargedens}$, the charge density $\chargedens$ relates to the
"bulk"-quantity $\boldsymbol{D}$, which does not incorporate ion
correlations.
	
Despite these differences, the resulting model equations are very
similar. This can be seen as follows. The complete set of equations
\cref{eq:non_dim_poisson_small_pots,eq:non_dim_forcelaw_small_pots}
for the limit of small potentials can be cast into one equation for
the electric potential alone,
\begin{equation}
   \dielrelop \tilde{\bn}^2\tilde{\elpot} = \tilde{\elpot},
\end{equation}
where the dielectric operator $\dielrelop$ is defined in
\cref{eq:dielectric_operator}. Noting the conceptual similarity
between the dielectric operators $\dielrelop$ and
$\hat{\bar{\varepsilon}}$, the similarity between our model and BSK
theory becomes apparent. In this way, we give physical meaning to the correlation length $\ell_{\ce{c}}$ in BSK theory and outline its calculation.

Finally, we emphasize that the higher-order gradient-terms, which are
phenomenologically incorporated in the BSK ap\-proach, emerge naturally
within our rigorous continuum mo\-del. In particular, they merely
constitutes the limiting case for small potentials of the more
fundamental integral formulation \cref{eq:non_dimm_intgrtd_intgrl_tt}. Furthermore, in
contrast to the phenomenological BSK model, our order-expansion
comprises also a zero-order correction in the dielectric operator, see
\cref{eq:dielectric_operator}. This mode is mandatory to realize the
``complete'' phase-space of interfacial profiles (see \cref*{sec:SItriv-order-expans,sec:SIline-order-expans} in the SI).

\subsection{Outlook}
\label{sec:model_extension}

In this section, we discuss how our framework can be extended to account for additional microscopic IL effects.

In this work, we have supplemented our bulk description for ILs and highly correlated electrolytes, recently presented in Ref. \citenum{schammer2020theory}, by non-local interactions. Furthermore, we applied the resulting framework for the case of short-ranged hardcore interactions. However, the generality of our framework based on the modelling of the free energy, offers the possibility to incorporate a wide range of non-local effects into our framework. 

This includes properties like ion asymmetry, ion geometry, polarization,
and charge delocalization, which have a significant influence on the structure of ILs near electrified interfaces.\cite{de2022structural,li2013effect,wei2021effects,izgorodina2009components,paek2015influence} These effects result partly from the relative orientation between the ions, which makes a one-dimensional approach challenging. 
Nevertheless, assuming a highly symmetric set-up, the one-dimensional description might still capture some basic consequences of these effects.

Similar to the force fields used in atomistic simulations, the short-ranged repulsive interaction can be supplemented by a longer-ranged attractive tail, taking account for higher-order electrostatic effects of van-der-Waals type, or for larger ions of complex geometry, \emph{i.e.}, long alkyl chains.\cite{bedrov2019molecular,heinz2013thermodynamically,refId0}  Also, by refining the short-ranged repulsive interaction potential, more detailed models for the ion geometry and ion asymmetry can be incorporated into our model. However, the strong influence of these microscopic properties on the EDL structure may lead to some novel features within our framework. For example, the three energy scales which determine the screening profile might transition into field quantities which exhibit spatial variation. Also, the phase space of screening profiles might become higher-dimensional, which can lead to a more complex set of phase boundaries allowing for "mixed" screening types.

Non-trivial polarization effects could be incorporated into our linear constitutive model for the coupling between the electric field and the dielectric displacement. This would result in a spatially varying dielectric function $\dielrel(\xcomp)$ appearing in the electrostatic energy scale $\Eel$ (see \cref{eq:en_scale_eel}) and a direct coupling of the chemical potentials with the ion polarization. For small dielectric perturbations, we hypothesis that the electrostatic energy scale gets more "diffuse", which has the effect that the phase boundaries between the screening profiles wash out. Only in the case of large dielectric variations, we expect the phase space of screening profiles to be altered significantly.

Our dynamical theory offers the possibility to  investigate transport processes occurring in electrochemical devices, \emph{e.g.}, the influence of EDL-charging on the electrolyte performance, or the influence of the EDL structure on the electrode transfer kinetics. However, electrochemical devices have some characteristic properties which must be carefully taken account for when they are modelled. For example, overlapping double layers in nano-porous electrodes could be taken account for.\cite{lian2021effects}

In our description we assumed an ideally flat surface, which can be a bad approximation for many electrochemical systems.\cite{jansch2015influence}. The influence of interface roughness on the EDL structure can be modelled by modifying the entropic contributions in the free energy functional.\citenum{aslyamov2021electrolyte} In our analysis, this would alter the thermal energy scale $\Eth$, see \cref{eq:en_scale_eth}. Depending upon the surface morphology, this would enhance the disordering effect of the thermal energy on the EDL structure. As result, the formation of crystalline phases might become suppressed at rougher surfaces, similar to increasing the temperature.

\section{Conclusion}
\label{sec:conclusion-1}
	
In this work, we complement our thermodynamically consistent continuum
framework for IL electrolytes by non-local molecular repulsion. Our
integral formulation can be determined by ab-initio MD
simulations. Assuming short-ranged interactions, we expand the
interaction free energy in gradients of the concentrations and adjust
the dynamic equations for transport. The resulting equations connect
to the phenomenologic approach of BSK theory. We validate our approach
by simulations and find a remarkable agreement between the different
variants of our theory.
	
In this way, we develop a predictive multi-scale approach to the
theory of ILs at electrified interfaces. Atomistic density functional
theory calculations parameterize MD simulations, MD simulations yield
an integral formulation for molecular repulsion in our thermodynamic
consistent transport theory, our theory can be expanded to give the
phenomenological BSK theory.
	
The expanded continuum approach allows to perform analytic asymptotic
analysis which creates deeper insights into parameter dependence of
EDL structure as we demonstrate for the example of binary ILs. First,
we have neglected molecular repulsion. We can analytically describe
both limits, the dilute Debye limit, where charge density is
exponentially decaying, and the concentrated crowding limit, where
charge is saturated due to steric effects. Second, we have taken into account molecular repulsion. We discuss the
structure of the EDL dependent on energy scales for thermal motion,
molecular repulsion, and electric Coulomb forces and find three
different phases. For small interactions, we recover the dilute Debye
limit. For intermediate interactions, a multi-layer structure of ions
emerges which is washed out over several atom layers. For very large
interactions, the analysis predicts a long-ranged, non-decaying
crystalline order of the EDL. In simulations of our full theory, we
eventually observe charge ordering of quasi-crystalline multi-layers
in this case.
	
In summary, we have proposed a thermodynamic consistent description of
ILs at electrified interfaces that closes a gap in their multi-scale
understanding. This makes possible a predictive theoretical approach
for tailoring ILs. We proof that the inter-molecular forces determine
the EDL structure of binary ILs. Future works should extend the work
to ternary mixtures of ILs and incorporate the shape of molecules into the theory.

\section*{Supporting Information}
Transport Theory of Interacting Electrolytes (including Gradient Expansion, Functional Derivative); Non-Dimensional Formulation; Binary Ionic Liquid (including Chemical Forces, Symmetric Ion Species, Charge Saturation, Charge Oscillations); Computational Details; Parameters Binary Ionic Liquid; Relation to Experimental Methods; Simulation Results

\section{Acknowledgment}
\label{sec:ackn-luzibmbf}

This work was supported by the European
Union's Horizon 2020 research and innovation program via the
``Si-DRIVE'' project (grant agreement No 814464).

The authors acknowledge support by the German Research Foun\-dation
(DFG) through grant no INST 40/575-1 FUGG (JUSTUS 2 cluster) and the
state of Baden-W\"urt\-tem\-berg through bwHPC.

\bibliographystyle{apsrev4-2}
\bibliography{bibliography}	

\end{document}